\documentclass[english,opre,nonblindrev]{informs3_arxiv}
\usepackage[T1]{fontenc}
\usepackage[latin9]{inputenc}
\usepackage{color}
\usepackage{array}
\usepackage{float}
\usepackage{booktabs}
\usepackage{varwidth}
\usepackage{amsmath}
\usepackage{amssymb}
\usepackage{graphicx}
\usepackage{geometry}
\usepackage{esint}
\usepackage[authoryear]{natbib}

\makeatletter

\providecommand{\tabularnewline}{\\}
\newenvironment{cellvarwidth}[1][t]
    {\begin{varwidth}[#1]{\linewidth}}
    {\@finalstrut\@arstrutbox\end{varwidth}}

\usepackage{amsmath,bbm}
\usepackage{footnote}
\DoubleSpacedXI % Made default 4/4/2014 at request
\usepackage{endnotes}
\let\footnote=\endnote

\usepackage{pgfplots}
\usepackage{natbib}
 \bibpunct[, ]{(}{)}{,}{a}{}{,}%
 \def\bibfont{\small}%
\newcommand{\I}{\mathbbm{1}}

\usepackage[none]{hyphenat}
\makeatletter

\usepackage{endnotes}
\let\footnote=\endnote

\TheoremsNumberedThrough     % Preferred (Theorem 1, Lemma 1, Theorem 2)
\ECRepeatTheorems

\EquationsNumberedThrough    % Default: (1), (2), ...

\def\footnoterule{\kern-3\p@
  \hrule \@width 6.5in \kern 2.6\p@} % the \hrule is .4pt high

\usepackage{algorithm}
\usepackage{algpseudocode}

\definecolor{blue}{RGB}{0,0,0}
\usepackage{xcolor}
\usepackage[linesnumbered,ruled,vlined,algo2e]{algorithm2e}
\usepackage{setspace}
\SetKwInput{KwInput}{Input}                % Set the Input
\SetKwInput{KwOutput}{Output}              % set the Output
\makeatother

\usepackage{babel}
\begin{document}
\title{Angular Combining of Forecasts of Probability Distributions}

\RUNAUTHOR{Taylor and Meng}
\RUNTITLE{Angular Combining}

\ARTICLEAUTHORS{% 
\AUTHOR{ James W. Taylor}
\AFF{Saïd Business School, University of Oxford, UK, \EMAIL{james.taylor@sbs.ox.ac.uk}} 
\AUTHOR{Xiaochun Meng}
\AFF{School of Management, University of Bath, UK, \EMAIL{xiaochun.meng@bath.edu}}
}

\abstract{When multiple forecasts are available for a probability distribution, forecast combining enables a pragmatic synthesis of the information to extract the wisdom of the crowd. The linear opinion pool has been widely used, whereby the combining is applied to the probabilities of the distributional forecasts. However, it has been argued that this will tend to deliver overdispersed distributions, prompting the combination to be applied, instead, to the quantiles of the distributional forecasts. Results from different applications are mixed, leaving it as an empirical question whether to combine probabilities or quantiles. In this paper, we present an alternative approach. Looking at the distributional forecasts, combining the probabilities can be viewed as vertical combining, with quantile combining seen as horizontal combining. Our proposal is to allow combining to take place on an angle between the extreme cases of vertical and horizontal combining. We term this angular combining. The angle is a parameter that can be optimized using a proper scoring rule. For implementation, we provide a pragmatic numerical approach and a simulation algorithm. Among our theoretical results, we show that, as with vertical and horizontal averaging, angular averaging results in a distribution with mean equal to the average of the means of the distributions that are being combined. We also show that angular averaging produces a distribution with lower variance than vertical averaging, and, under certain assumptions, greater variance than horizontal averaging. We provide empirical results for distributional forecasts of Covid mortality, macroeconomic survey data, and electricity prices.}


\maketitle

\section{Introduction\label{Section 1}}

Predictions of probability distributions convey forecast uncertainty
and are of great importance for many applications. \textcolor{blue}{For
example, }forecasts of a distribution\textquoteright s tails are the
focus in economic and financial risk management (\citealt{Brownlees2021};
\citealt{Wang2021}); quantile forecasts are needed to enable optimal
decision making in newsvendor contexts (\textcolor{blue}{\citealt{Arrow1951}});
efficient call center workforce planning requires forecasts of the
distribution and quantiles of call arrival rates (\citealt{Ye2019});
\textcolor{blue}{and, }in epidemic modeling, distributional forecasts
provide situational awareness, sometimes in the form of interval forecasts,
to support health policy \textcolor{blue}{decisions} (\citealt{Bracher2021}). 

When forecasts are available from multiple experts or methods, combining
is often used as a convenient way to capture the wisdom of the crowd
through the synthesis of the different sources of forecast information.
\citet{winkler2019probability} provide a review of the growing literature
on approaches used to combine probabilistic forecasts. They comment
that averaging has the appeal, not only of simplicity, but also robustness
and competitive empirical performance. However, once a reasonable
history of past accuracy becomes available, weighted combinations
can be more accurate (\citealt{Aastveit2019}). When a reasonably
large number of distributional forecasts are available, the combining
method could be chosen as the median (see \citealt{Hora2013}) or
an approach based on trimmed means. Quite apart from enabling robustness
to outliers, trimming the outermost or innermost distributions from
a set of distributional forecasts can be used to control the dispersion
of the resulting combined distributional forecast (\citealt{Jose2014};
\citealt{GrushkaCockayne2017}). 

To combine forecasts of continuous distributions, \citet{lichtendahl2013better}
consider whether it is better to average probabilities or quantiles.
They note that, although averaging probabilities is more common, it
is a concern that, for a set of calibrated distributional forecasts,
averaging probabilities will deliver a forecast that is not calibrated
and is too wide (see also \citealt{Hora2004}; \citealt{gneiting2013combining}).
\citet{lichtendahl2013better} point out that averaging quantiles
has some appealing theoretical properties, most notably that it leads
to a distributional forecast \textcolor{blue}{that has lower variance
than} averaging probabilities. However, individual distributional
forecasts can be underdispersed, in which case averaging quantiles
can produce an underdispersed distributional forecast. Indeed, drawing
on empirical results, \citet{Cooke2021} advise against averaging
quantiles. Given conflicting recommendations, a pragmatic approach
is to consider it an empirical question as to whether it is better,
for a particular application, to combine probabilities or quantiles.
Ideally, one should let the data decide.

In this paper, for continuous distributions, we propose an alternative
averaging approach that lies between averaging probabilities and averaging
quantiles. With averaging probabilities viewed as averaging vertically,
and averaging quantiles viewed as averaging horizontally, our alternative
is to average at an angle between these two extremes. The angle is
a parameter that can be optimized using past data and a proper scoring
rule. With the angle suitably chosen, the new method can be viewed
as finding the right balance between vertical and horizontal averaging.
This \textit{angular averaging} is a practical proposal that extends
and encompasses vertical and horizontal averaging. 

Our theoretical results build on several results provided by \citet{lichtendahl2013better}
for vertical and horizontal averaging. We show that, like vertical
and horizontal averaging, angular averaging delivers a distribution
for which the mean is the average of the means of the distributions
in the combination. We also show that angular averaging produces a
distribution with lower variance than vertical averaging, and, under
certain assumptions, greater variance than horizontal averaging. \textcolor{blue}{Furthermore,
under these assumptions, the variance is a monotonic function of the
angle. }Drawing on our results for the mean and variance of angular
averaging, we address when to use this combining method by describing
a situation where it should be preferred to horizontal and vertical
averaging. For a popular scoring rule, we find that the score for
angular averaging is no worse than the average score of the distributional
forecasts in the combination. Other forms of angular combining can
also be considered, such as performance-based weighted averaging.
\citet{Hora2013} show that median aggregation results in the same
distribution regardless of whether the median is obtained vertically
or horizontally. We find that the same distribution is produced when
the median is obtained at an angle. 

Section~\ref{Section 2} provides background for our proposal by
discussing vertical and horizontal combining. Section~\ref{Section 3}
presents angular averaging followed by consideration of other forms
of angular combining. Section~\ref{Section 4} presents theoretical
results for angular combining. Section~\ref{Section 5} provides
our main empirical study, which evaluates the new proposals using
forecasts for weekly Covid mortality at the national and state level
in the U.S. Section~\ref{Section 6} presents additional empirical
studies, involving forecasts of macroeconomic variables and electricity
prices. Section~\ref{Section 7} provides a summary and concluding
comments.

\section{Vertical and Horizontal Combining of Distributional Forecasts\label{Section 2}}

In this section, we describe the contrasting approaches of vertical
and horizontal combining. We go into some depth, as we feel this is
important for the presentation of our new proposals in Section~\ref{Section 3}.
For simplicity, we focus mainly on averaging, and then briefly consider
other combining approaches. 

\subsection{Vertical and Horizontal Averaging\label{Section 2.1}}

\begin{figure}[H]
\begin{centering}
\caption{\linespread{0.50}\selectfont{}\label{fig: Vertical averaging}Vertical
averaging (in black) of two Gaussian CDFs (in grey) in (a), and corresponding
PDFs in (b).}
\vspace{-0.2cm}
\par\end{centering}
\centering{}\includegraphics[scale=0.65]{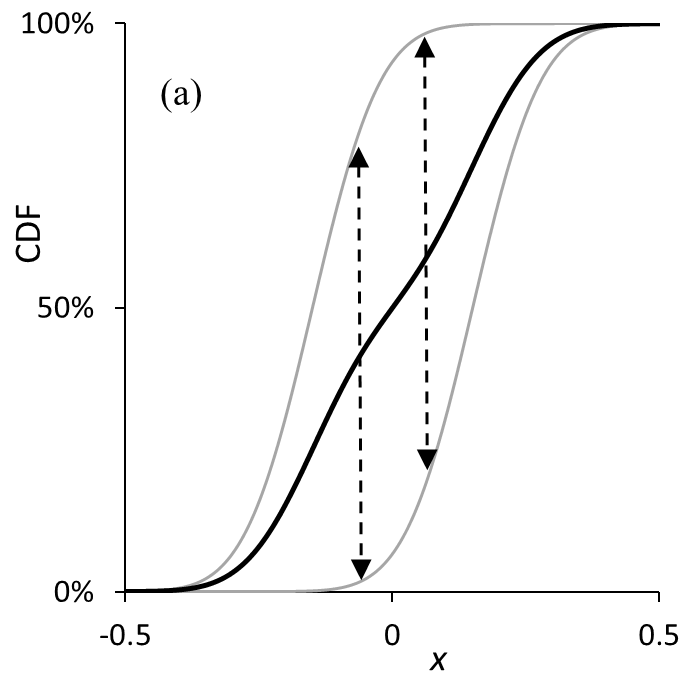}\includegraphics[scale=0.65]{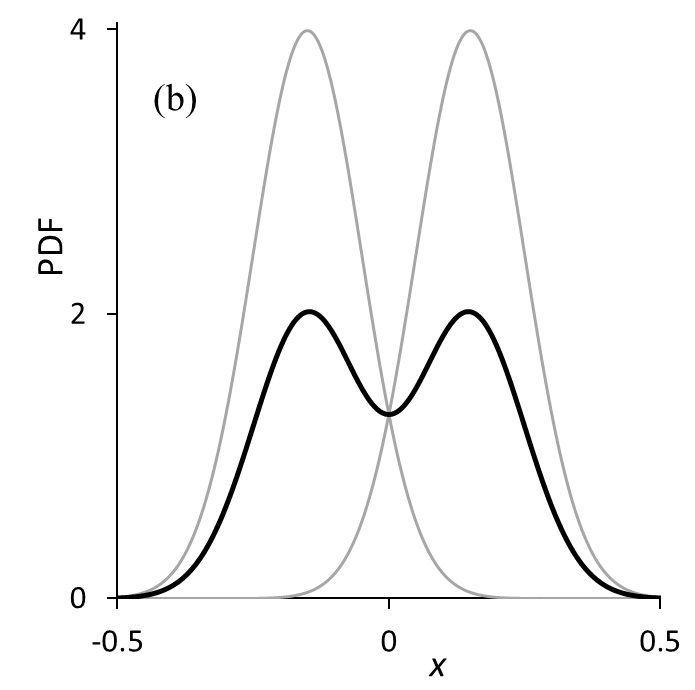}\vspace{-0.8cm}
\end{figure}

\begin{figure}[H]
\begin{centering}
\caption{\linespread{0.50}\selectfont{}\label{Fig: Horizontal averaging}
Horizontal averaging (in black) of two Gaussian CDFs (in grey) in
(a), and corresponding PDFs in (b).}
\vspace{-0.2cm}
\par\end{centering}
\centering{}\includegraphics[scale=0.65]{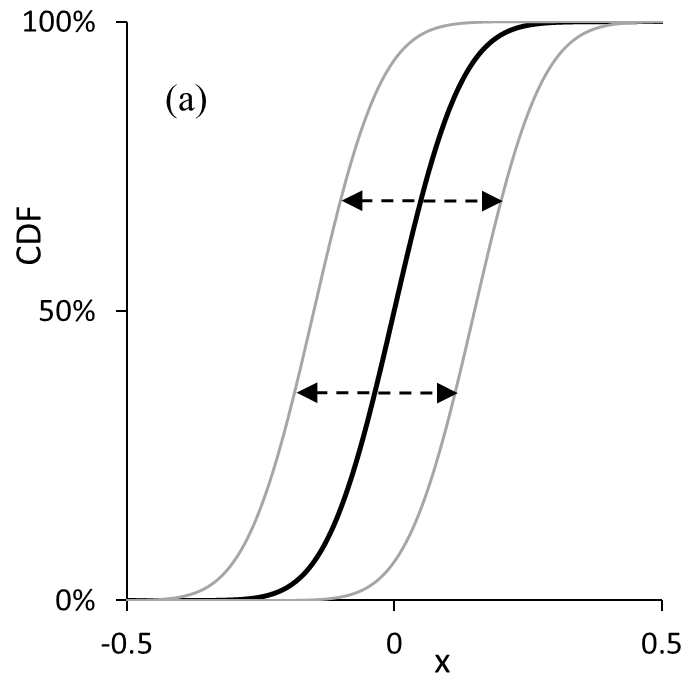}\includegraphics[scale=0.65]{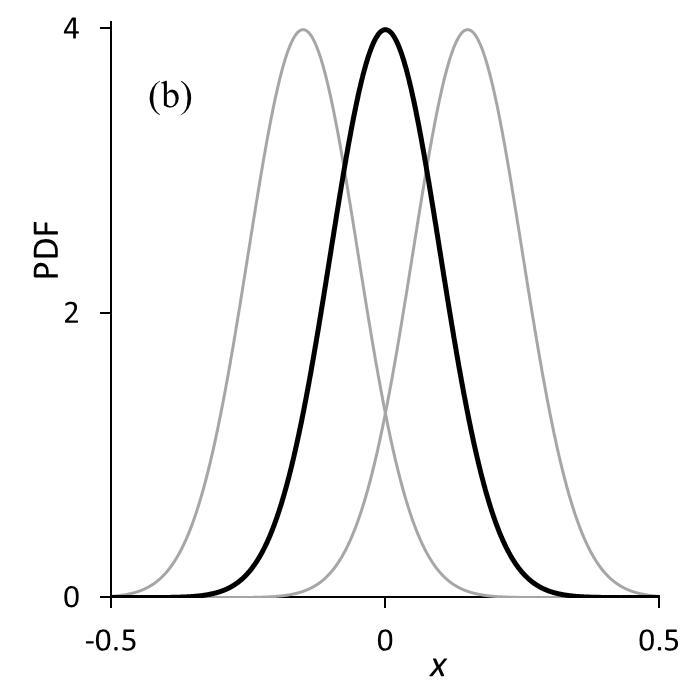}\vspace{-0.8cm}
\end{figure}

\citet{lichtendahl2013better} illustrate the differences that can
result from vertical and horizontal averaging using figures showing
the averaging of two Gaussian cumulative distribution functions (CDFs)
with different means but identical variances. We repeat similar figures
here to assist the explanation of our proposals later in the paper.
Figure~\ref{fig: Vertical averaging}(a) shows the CDF that results
from the vertical averaging of two Gaussian CDFs that have means of
-0.15 and 0.15, and standard deviations of 0.1. Figure~\ref{fig: Vertical averaging}(b)
presents the corresponding probability density functions (PDFs). Figure~\ref{Fig: Horizontal averaging}(a)
shows the CDF resulting from horizontal averaging of the two Gaussian
CDFs. The corresponding PDFs are presented in Figure~\ref{Fig: Horizontal averaging}(b).
Comparing Figures~\ref{fig: Vertical averaging} and \ref{Fig: Horizontal averaging},
we see that noticeably different distributions can result from the
two forms of averaging. Figure~\ref{fig: Vertical averaging} shows
that a bimodal CDF has resulted from the vertical averaging. By contrast,
Figure~\ref{Fig: Horizontal averaging} illustrates the result that
if the individual distributions are from the same location scale family,
such as Gaussian, horizontal averaging leads to a distribution from
the same family (\citealt{Thomas1980}).

We note that \textcolor{blue}{horizontal} combining cannot be used
if part of an individual CDF forecast is \textcolor{blue}{horizontal},
because, for the corresponding value on the \textcolor{blue}{$y$}-axis,
there would be a set of values of the CDF, and so it would not be
clear what CDF value to include in the combination. Therefore, in
our discussions of horizontal and vertical combining, we assume each
CDF forecast is strictly monotonic. 

In this paper, we consider the forecasting of continuous random variables.
If we write the CDF and PDF of an individual forecast distribution
as $F_{i}$ and $f_{i}$, respectively, the following provides the
CDF $F_{V}$ and PDF $f_{V}$ of the vertical average of $k$ individual
forecast distributions, 
\begin{align}
F_{V}(x) & =\frac{1}{k}\sum_{i=1}^{k}F_{i}(x)\quad\textrm{and}\quad f_{V}(x)=\frac{1}{k}\sum_{i=1}^{k}f_{i}(x).\label{eq: Vertical averaging CDF =000026 PDF}
\end{align}
The PDF $f_{V}$ can be formally derived by differentiating the CDF
with respect to (w.r.t.) $x$. 

Given that horizontal averaging involves averaging quantiles, and
the $\alpha$ quantile of an individual forecast distribution is $F_{i}^{-1}(\alpha)$,
the first of the following two equations provides an expression in
terms of the CDF $F_{H}$ resulting from horizontal averaging, 
\begin{equation}
F_{H}^{-1}(\alpha)=\frac{1}{k}\sum_{i=1}^{k}F_{i}^{-1}(\alpha)\quad\textrm{and}\quad f_{H}\left((F_{H}^{-1}(\alpha)\right)=\frac{1}{\frac{1}{k}\sum_{i=1}^{k}1/f_{i}\left(F_{i}^{-1}(\alpha)\right)}.\label{eq: Horizontal averaging CDF =000026 PDF}
\end{equation}
\citet{Bamber2016} show that differentiating w.r.t. $\alpha$ each
side of the first equation in (\ref{eq: Horizontal averaging CDF =000026 PDF})
gives the second equation, which provides an expression for the PDF
$f_{H}$ produced by horizontal averaging. 

\subsection{Is it better to Average Vertically or Horizontally?\label{Section 2.2}}

Figures~\ref{fig: Vertical averaging} and \ref{Fig: Horizontal averaging}
illustrate the point, shown theoretically by \citet{lichtendahl2013better},
that vertical averaging produces a CDF with greater variance than
horizontal averaging. They suggest that the CDF forecast produced
by vertical averaging will often be too wide, prompting them to propose
horizontal averaging as an alternative. Using CDF forecasts from surveys
of professional macroeconomic forecasters, they provide empirical
results supporting the use of horizontal averaging.

With other applications, it is not clear whether it is preferable
to combine vertically or horizontally. For electricity price forecasting,
\citet{Uniejewski2019} and \citet{Marcjasz2020} find that vertical
averaging is more accurate, while \citet{Uniejewski2021} obtain better
results for horizontal averaging, with vertical averaging leading
to CDFs that are too wide. For forecasting wave height, \citet{taylor2017probabilistic}
report that vertical and horizontal averaging deliver similar accuracy.
The fact that horizontal averaging delivers a CDF with lower variance
than vertical averaging leads \citet{Bamber2016}, \citet{Colson2017}
and \citet{Cooke2022} to express concern that horizontal averaging
can result in a distribution that is too narrow. Using data from multiple
studies in a variety of applications involving expert judgments of
probability distributions, they find vertical averaging noticeably
more accurate than horizontal averaging. 

In some applications, only a set of quantile forecasts are available
(see, for example, \citealt{Makridakis2022}). Indeed, this is the
case with our Covid dataset. In such situations, \citet{Cooke2021}
and \citet{Cooke2022} acknowledge that horizontal averaging is easier,
as vertical averaging requires a method to impute the complete CDF
forecasts from the quantile predictions. Despite this, they warn against
horizontal averaging, citing its tendency to lead to CDFs that underestimate
the uncertainty. 

\subsection{Other Forms of Vertical and Horizontal Combining\label{Section 2.3}}

Instead of using the simple average, distributional forecasts are
often combined using weighted linear combinations. The vertical weighted
combination $F_{V,w}$ and horizontal weighted combination $F_{H,w}$
are expressed as follows, 
\begin{align*}
F_{V,w}(x) & =\sum_{i=1}^{k}w_{i}F_{i}(x)\quad\textrm{and}\quad F_{H,w}^{-1}(\alpha)=\sum_{i=1}^{k}w_{i}F_{i}^{-1}(\alpha)
\end{align*}
where the $w_{i}$ are the combining weights. To ensure that the result
is a well-defined CDF, vertical combining requires convex weights,
i.e., $w_{i}\geq0$ and $\sum_{i=1}^{k}w_{i}=1$, while horizontal
combining requires only that the weights are nonnegative. It is natural
for the weights to reflect relative accuracy. For example, they can
be optimized with the objective function set as a scoring rule, such
as the log score or continuous ranked probability score (CRPS) (see,
for example, \citealt{hall2007combining}; \citealt{Raftery2005}).
The vertical weighted combination is sometimes termed the linear opinion
pool (\citealt{Stone1961})\textcolor{blue}{,} and has been described
as the most popular approach to combining distributional forecasts
(\citealt{Knueppel2022}), with equal weighting viewed as a special
case. Using weights in proportion to the log score of each forecast,
\citet{Busetti2017} finds that horizontal combining is more accurate
than vertical combining for quarterly economic growth and inflation.
In a weighted horizontal combining method, \citet{Taylor2023} set
the weights to be inversely proportional to an approximation to the
CRPS. If a different weighting scheme is considered desirable for
different parts of the distribution (see, for example, \citealt{kapetanios2015generalised};
\citealt{fakoor2023flexible}), restrictions or adjustments would
be needed to avoid quantile crossing (\citealt{chernozhukov2010quantile}). 

The tendency for vertical combining to deliver overdispersed distributional
forecasts has motivated nonlinear forms of vertical combining. The
logarithmic opinion pool, which can be viewed as a geometric weighted
average of PDFs, has been considered (see, for example, \citealt{Busetti2017};
\citealt{Cooke2022}). Building on this, \citet{gneiting2013combining}
introduce generalized linear combinations, which involve a linear
combination of CDF forecasts that have each been transformed using
a link function. They also propose a beta-transformed linear pool
that enables the combined CDF forecast to have a flexible variance,
controlled by optimizing the parameters of the beta transformation.
\citet{Meer2024} introduce CRPS-based online learning for this form
of combining.

\citet{Jose2014} use trimming to control the variance of the CDF
resulting from vertical averaging. They suggest entire CDF forecasts
can be trimmed based on the mean, or the trimming can be performed
separately, in a vertical sense, for different values of the outcome
variable. After trimming, the remaining CDF forecasts are vertically
averaged. Their exterior trimming approach reduces the impact of outlying
CDFs to reduce the variance of the resulting CDF, while their interior
trimming approach has the opposite effect by reducing the impact of
central lying CDFs. \citet{GrushkaCockayne2017} apply these ideas
to the CDFs produced in ensemble learning. 

\section{Angular Combining of Distributional Forecasts\label{Section 3}}

\subsection{Angular Averaging\label{Section 3.1}}

With arguments in favor of both vertical and horizontal averaging,
it becomes an empirical question as to which of the two approaches
should be chosen for any given application. However, it could well
be that neither vertical or horizontal averaging is ideal, and that
there are useful aspects of each approach. For example, the unknown
true density may be represented well by the tails produced by horizontal
averaging and the central part of the distribution produced by vertical
averaging.

\begin{figure}[H]
\begin{centering}
\caption{\linespread{0.50}\selectfont{}\label{fig: Angular averaging basics}Angular,
horizontal and vertical averaging of points on two Gaussian CDFs in
(a), and illustration of our parameterization in (b).}
\vspace{-0.2cm}
\par\end{centering}
\includegraphics[scale=0.65]{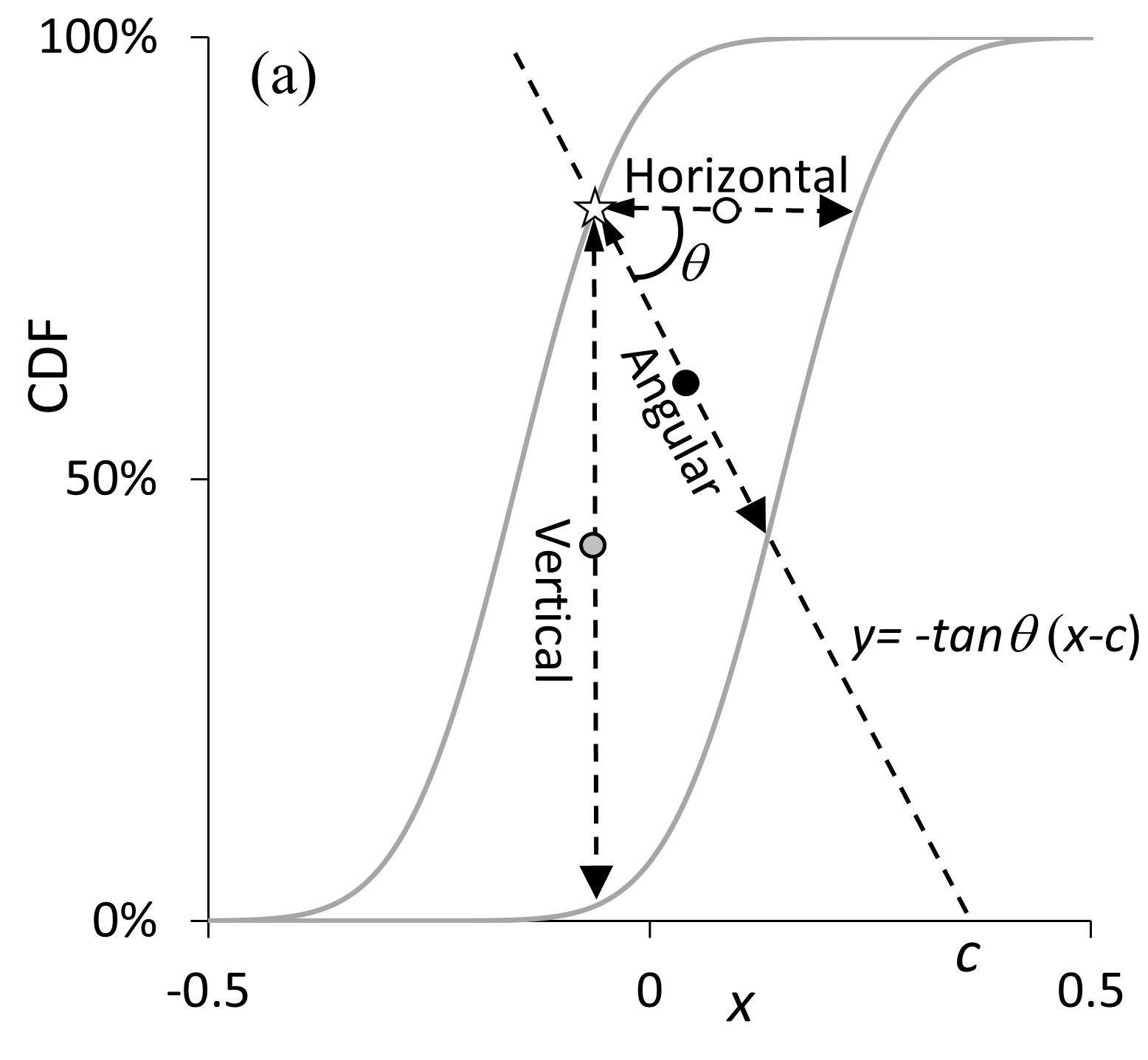}\includegraphics[scale=0.65]{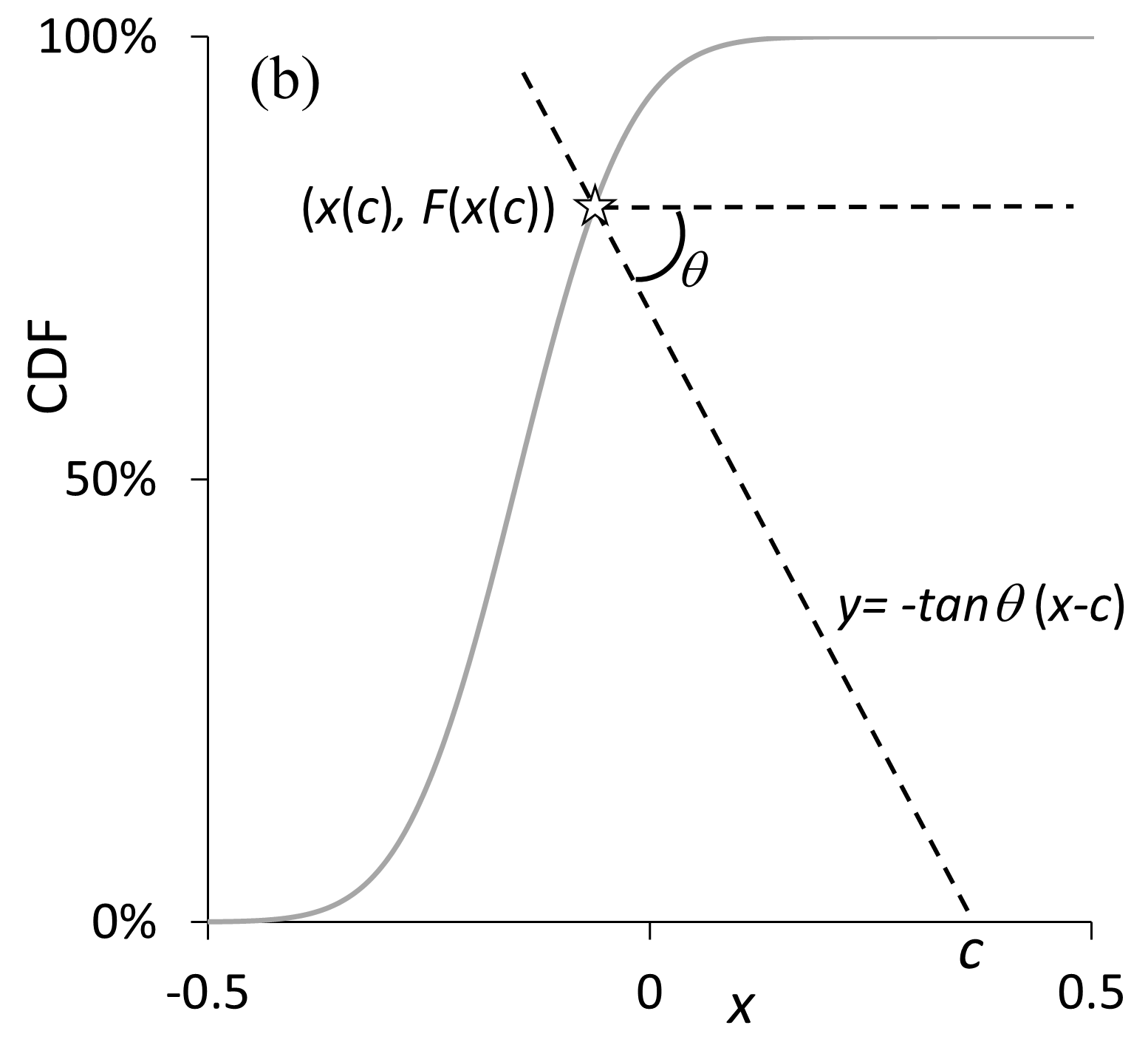}
\centering{}\vspace{-0.8cm}
\end{figure}

In this paper, we propose an alternative averaging approach that allows
the user to consider a range of possibilities between horizontal and
vertical averaging. Our pragmatic alternative is to average on an
angle that lies between these two extremes. We term this \textit{angular
averaging}. Figure~\ref{fig: Angular averaging basics}(a) illustrates
the idea with averaging of the same two CDFs considered in Figures~\ref{fig: Vertical averaging}
and \ref{Fig: Horizontal averaging}. Figure~\ref{fig: Angular averaging basics}(a)
shows angular averaging performed on a line that has been rotated
clockwise by an angle $\theta$ from the horizontal. Horizontal and
vertical averaging are special cases of angular averaging, corresponding
to $\theta$ being $0^{\textrm{o}}$ and $90^{\textrm{o}}$, respectively.
Figure~\ref{fig: Angular averaging basics}(a) shows horizontal,
vertical and angular averaging performed from the point indicated
by the white star on the left CDF. The three different forms of averaging
each lead to averaging of this point with a different point on the
right CDF. The result is three different averages, shown by the white-filled,
grey and black circular points.

Figure~\ref{fig: Angular averaging basics}(b) illustrates our parameterization
for angular averaging. For a fixed $\theta\in(0^{\textrm{o}},90^{\textrm{o}})$,
consider a family of straight lines $y=-\tan\theta(x-c)$, where $-\tan\theta$
and $c\in(-\infty,\infty)$ are the gradient and $x$-intercept, respectively.
We use $c$ to parameterize each CDF, and to enable a formal description
of the CDF resulting from angular averaging. For any given CDF $F$,
we define the point of intersection between the CDF curve $y=F(x)$
and the line $y=-\tan\theta(x-c)$ as $\left(x(c),F(x(c))\right)$.
The function $x(c)$ is increasing w.r.t. $c$, and $\left(x(c),F(x(c))\right)$
covers the entire CDF as $c$ ranges from $-\infty$ to $\infty$.
Thus, the set of points $\left(x(c),F(x(c))\right)$ provides a parameterization
of the CDF using $c$. That is, each point on the CDF can be uniquely
identified by a single value of $c$. Indeed, $F(x(c))$ is a function
of $c$ that behaves as a CDF. 

Using this parameterization for each of \textit{k} CDFs, $F_{i}$,
we have \textit{k} points of intersection $\left(x_{i}(c),F_{i}(x_{i}(c))\right)$
between the CDFs and the line $y=-\tan\theta(x-c)$. Averaging the
$k$ points $\left(\frac{1}{k}\sum_{i=1}^{k}x_{i}(c),\frac{1}{k}\sum_{i=1}^{k}F_{i}(x_{i}(c))\right)$
yields the CDF $F_{A,\theta}$ of the angular average at $\frac{1}{k}\sum_{i=1}^{k}x_{i}(c)$,\vspace{-0.6cm}

\begin{equation}
F_{A,\theta}\left(\frac{1}{k}\sum_{i=1}^{k}x_{i}(c)\right)=\frac{1}{k}\sum_{i=1}^{k}F_{i}(x_{i}(c))\label{eq: Angular averaging CDF}
\end{equation}
We note that (\ref{eq: Angular averaging CDF}) contains $\frac{1}{k}\sum_{i=1}^{k}x_{i}(c)$
and $\frac{1}{k}\sum_{i=1}^{k}F_{i}(x_{i}(c))$, which represent horizontal
and vertical averaging, respectively, implying that both of these
forms of averaging are present in angular averaging. The PDF $f_{A,\theta}$
of the angular average is provided by Proposition~\ref{prop: PDF_f_Atheta}.
The proof of this proposition is presented in Online Appendix A. If
$\theta=0$, $f_{A,\theta}$ reduces to harmonic averaging of the
PDFs, as in (\ref{eq: Horizontal averaging CDF =000026 PDF}) for
horizontal averaging. As $\theta$ increases and approaches $90^{\textrm{o}}$,
implying all the $x_{i}(c)$ take the same value, $f_{A,\theta}$
converges to the arithmetic average of the PDFs, as in (\ref{eq: Vertical averaging CDF =000026 PDF})
for vertical averaging.

\begin{proposition}\label{prop: PDF_f_Atheta}$f_{A,\theta}\left(\frac{1}{k}\sum_{i=1}^{k}x_{i}(c)\right)=\left(\sum_{i=1}^{k}\frac{f_{i}\left(x_{i}(c)\right)}{f_{i}\left(x_{i}(c)\right)+\tan\theta}\right)\Big/\left(\sum_{i=1}^{k}\frac{1}{f_{i}\left(x_{i}(c)\right)+\tan\theta}\right).$\end{proposition}

We also note that, although our parameterization is w.r.t. the $x$-intercept
$c$ of the angled line, we could have chosen to parameterize w.r.t.
the $y$-intercept of this line. Indeed, this would be needed for
horizontal averaging, as there is no $x$-intercept when the angled
line is horizontal.

We have two comments to make regarding the angle $\theta$. Firstly,
if and only if $\theta$ is an angle in the range $\left[0^{\textrm{o}},90^{\textrm{o}}\right]$,
will the function resulting from angular averaging be guaranteed to
be monotonic and take values in the range {[}0,1{]}, which are the
necessary conditions for a CDF. If $\theta$ was not a value in the
range $\left[0^{\textrm{o}},90^{\textrm{o}}\right]$, the angled line
in Figure~\ref{fig: Angular averaging basics} could be crossed more
than once by one of the CDFs to be combined. A second comment regarding
the angle $\theta$ is that it is a parameter that should, ideally,
be estimated using historical data. For many applications, past data
is available, and so the estimation of a single parameter is not particularly
burdensome. Indeed, the need for parameter estimation is also an aspect
of many other combining methods. Furthermore, even though horizontal
and vertical averaging require no parameter estimation, to decide
which of these two methods to use, there is a need for a sample of
past data with which to compare historical accuracy.

\begin{figure}[H]
\centering{}\caption{\linespread{0.50}\selectfont{}\label{fig: Angular averaging CDF =000026 PDF}Angular
combining (in black) of two Gaussian CDFs (in grey) in (a), and corresponding
PDFs in (b). Averaging is performed at an angle $\theta=45^{\textrm{o}}$,
as indicated by the dashed lines.}
\vspace{-0.2cm}
\includegraphics[scale=0.65]{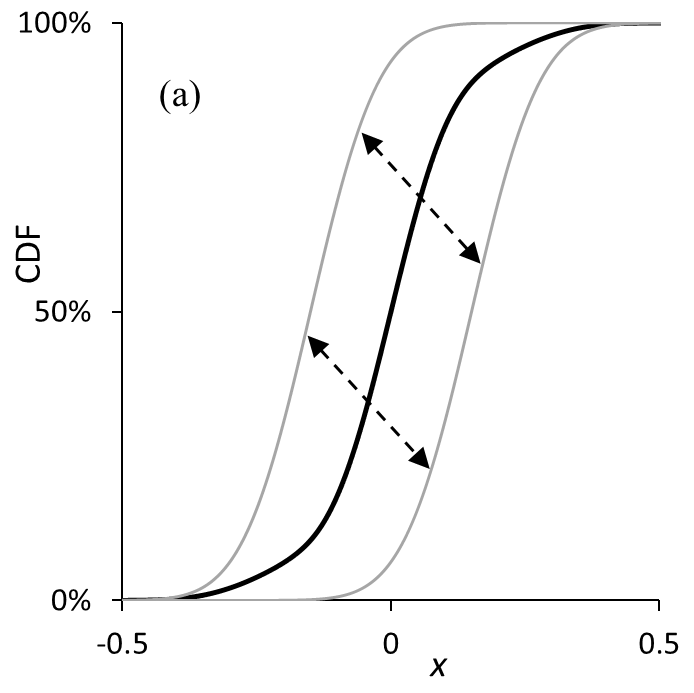}\includegraphics[scale=0.65]{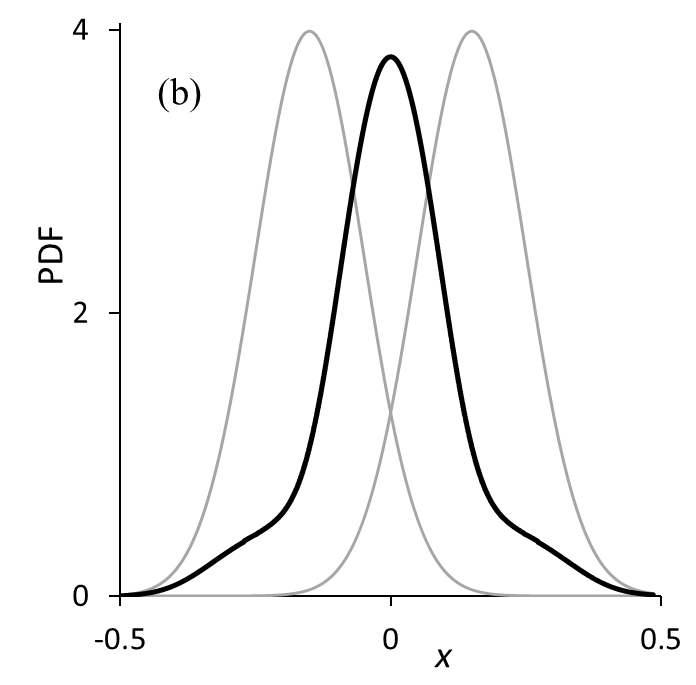}\vspace{-0.8cm}
\end{figure}

\begin{figure}[H]
\centering{}\caption{\linespread{0.50}\selectfont{}\label{fig: Angular averaging CDF =000026 PDF for different angles}Horizontal,
vertical and angular averaging of two Gaussian CDFs. Angular averaging
performed for $\theta=45^{\textrm{o}}$, $70^{\textrm{o}}$, $85^{\textrm{o}}$
and $88^{\textrm{o}}$. CDFs in (a) and PDFs in (b).}
\vspace{-0.2cm}
\includegraphics[scale=0.65]{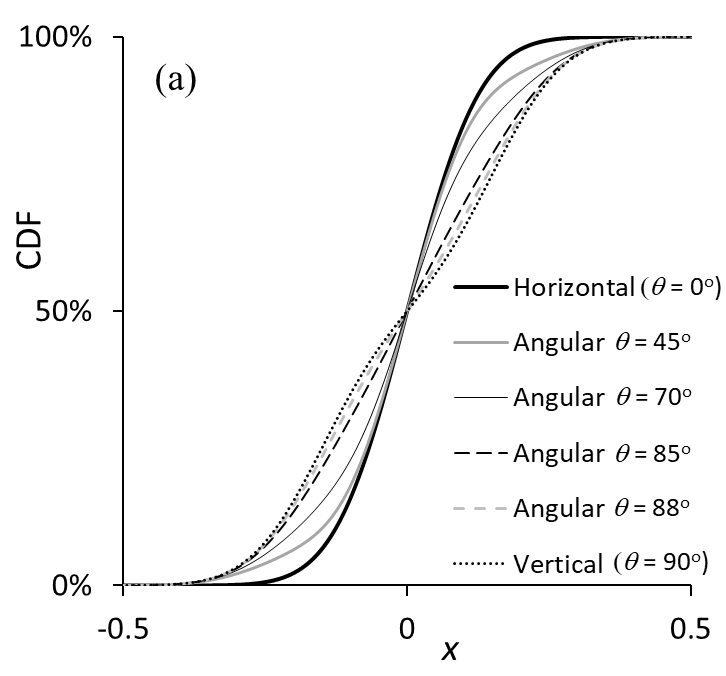}\includegraphics[scale=0.65]{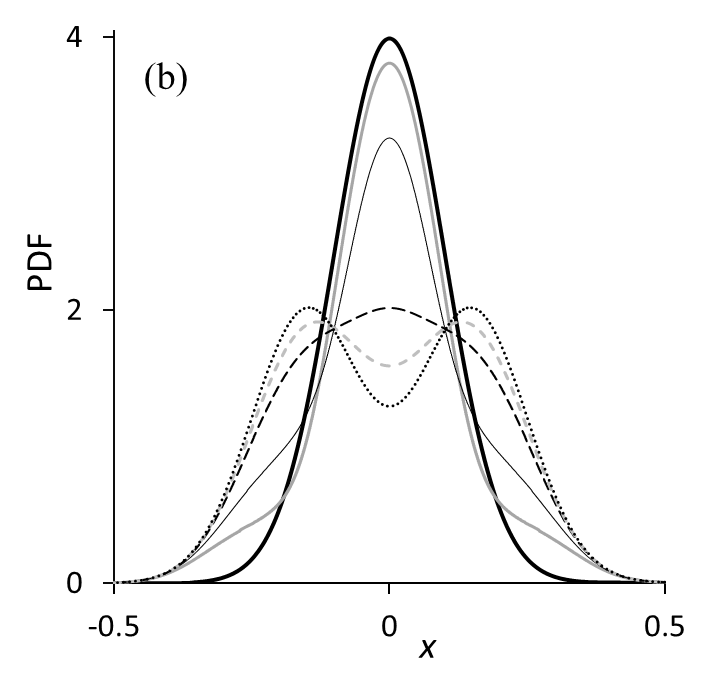}\vspace{-0.8cm}
\end{figure}

Figure~\ref{fig: Angular averaging CDF =000026 PDF} shows the CDF,
and corresponding PDF, resulting from angular averaging of the same
two Gaussian CDFs considered in Figures~\ref{fig: Vertical averaging}
to \ref{fig: Angular averaging basics}. The averaging has been performed
at an angle of $45^{\textrm{o}}$, as conveyed by the dashed lines
in Figure~\ref{fig: Angular averaging CDF =000026 PDF}(a). For a
variety of different values of $\theta$, Figure~\ref{fig: Angular averaging CDF =000026 PDF for different angles}(a)
shows the different CDFs that result from angular averaging of the
two Gaussian CDFs. Figure~\ref{fig: Angular averaging CDF =000026 PDF for different angles}(b)
provides the corresponding PDFs resulting from angular averaging.
As we noted when discussing Figures \ref{fig: Vertical averaging}(b)
and \ref{Fig: Horizontal averaging}(b), the PDFs resulting from vertical
and horizontal averaging of the two Gaussian CDFs are bimodal and
unimodal, respectively. It is interesting to note that, for the example
we are using, it is only when $\theta$ exceeds about $85^{\textrm{o}}$
that the angular averaging of the two CDFs leads to a PDF that is
bimodal. This suggests that it would be wise in empirical studies
to implement angular averaging for different values of $\theta$ spanning
the full range from $0^{\textrm{o}}$ to $90^{\textrm{o}}$. In our
empirical studies of Sections~\ref{Section 5} and \ref{Section 6},
as candidate values for $\theta$, we considered each integer value
from $0^{\textrm{o}}$ to $90^{\textrm{o}}$.

In Section~\ref{Section 2.1}, we commented that horizontal combining
cannot be applied if any part of an individual CDF forecast is horizontal.
We note that this is not a limitation of angular combining. Indeed,
for $0^{\textrm{o}}<\theta<90^{\textrm{o}}$, angular combining can
be applied to individual CDF forecasts of any type, provided each
is a well-defined CDF. 

\subsection{Other Forms of Angular Combining\label{Section 3.2}}

In comparing horizontal, vertical and angular combining, we have focused
on the simple average. However, CDFs have been combined horizontally
and vertically using other approaches, as we discussed in Section~\ref{Section 2.3}.
We suggest that such methods can be applied on an angle. As an example,
in our empirical study of Section~\ref{Section 5}, we consider a
combining method that weights forecasts inversely proportional to
the CRPS, and implement this horizontally, vertically and at angles
between $0^{\textrm{o}}$ and $90^{\textrm{o}}$. Using the parameterization
of Section~\ref{Section 3.1}, angular weighted combining is written
as
\begin{align}
F_{A,w,\theta}\left(\sum_{i=1}^{k}w_{i}x_{i}(c)\right) & =\sum_{i=1}^{k}w_{i}F_{i}(x_{i}(c)).\label{eq: Angular wtd combining CDF}
\end{align}
\vspace{-0.9cm}

\subsection{Angular Combining as Generalized Linear Combining\label{Section 3.3}}

\begin{figure}[H]
\begin{centering}
\caption{\linespread{0.50}\selectfont{}\label{fig: Angular as generalized linear combining}Visual
illustration of angular averaging as generalized linear averaging.
The light grey point is the angular average of the points of intersection
of the angled line with the two CDFs, $F_{1}$ and $F_{2}$. The transformation
of this light grey point results in the black point, which is also
the vertical average of the two transformed CDFs, $h_{\theta}(F_{1})$
and $h_{\theta}(F_{2})$, at the point $x=c$.}
\vspace{-0.2cm}
\par\end{centering}
\centering{}\includegraphics[scale=0.65]{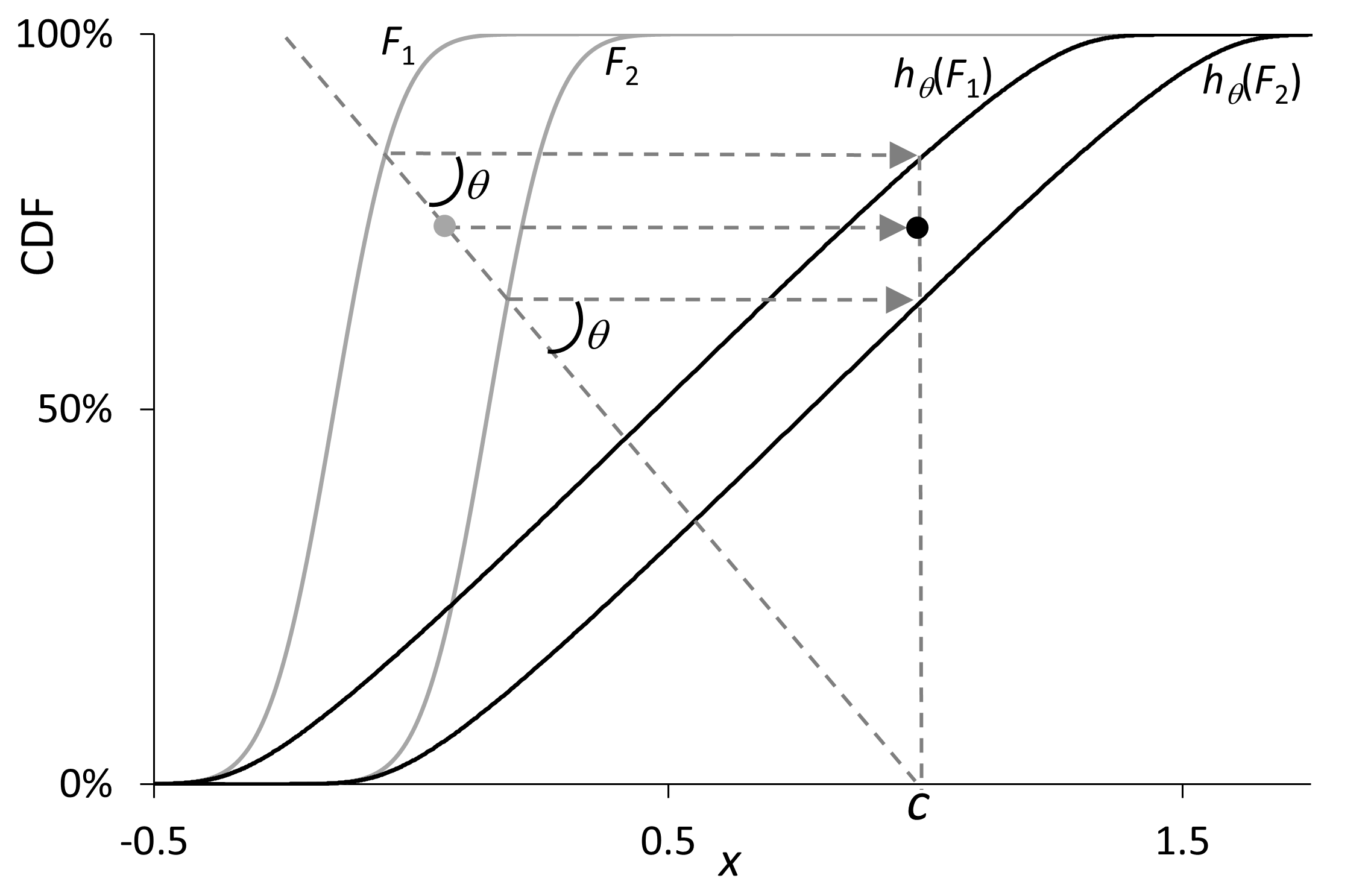}
\end{figure}

A useful way to understand angular combining is through the framework
of the generalized linear combination in (\ref{eq: Generalised linear combining}),
considered by \citet{gneiting2013combining},
\begin{equation}
h\left(F_{GL,w,h}(x)\right)=\sum_{i=1}^{k}w_{i}h(F_{i}(x))\label{eq: Generalised linear combining}
\end{equation}
where $h$ is a continuous and strictly increasing link function.
In simple terms, the transformed CDF $F_{GL,w,h}$ under $h$ is the
vertical combination of the transformed $F_{i}$ under $h$. Angular
combining can be viewed as similar in spirit, but with the notable
difference that a particular link function is used, and it is applied
to the quantiles instead of the CDFs. We illustrate this in Figure~\ref{fig: Angular as generalized linear combining}
for the angular averaging of two CDFs in light grey, $F_{1}$ and
$F_{2}$. The figure shows $F_{1}$ and $F_{2}$ transformed horizontally
to give two new CDFs in dark grey, $h_{\theta}(F_{1})$ and $h_{\theta}(F_{2})$,
using the link function $h_{\theta}$ in (\ref{eq: Quantile transformation link function}),
which defines a quantile transformation,
\begin{equation}
h_{\theta}(F)^{-1}(\alpha)=F^{-1}(\alpha)+\frac{\alpha}{\tan\theta}\quad\quad\forall\alpha\in[0,1].\label{eq: Quantile transformation link function}
\end{equation}
In Figure~\ref{fig: Angular as generalized linear combining}, the
light grey point on the angled line is the angular average of the
points of intersection of the angled line with $F_{1}$ and $F_{2}$.
The transformation of this light grey point results in the black point,
which is also the vertical average of $h_{\theta}(F_{1})$ and $h_{\theta}(F_{2})$
at $x=c$, where $c$ is the $x$-intercept of the angled line, as
defined in Section~\ref{Section 3.1}. Based on this, we can reinterpret
the angular combining in (\ref{eq: Angular wtd combining CDF}) within
a framework similar to (\ref{eq: Generalised linear combining}),
\begin{equation}
h_{\theta}\left(F_{A,w,\theta}\right)(c)=\sum_{i=1}^{k}w_{i}h_{\theta}(F_{i})(c).\label{eq: Angular as generalized linear combining}
\end{equation}

In Online Appendix B, we discuss how (\ref{eq: Quantile transformation link function})
and (\ref{eq: Angular as generalized linear combining}) allow us
to view angular combinations, like vertical and horizontal combinations,
as barycenters of the set of individual CDF forecasts.

\vspace{-1.2cm}

\subsection{Practical Implementation of Angular Combining\label{Section 3.4}}

\begin{figure}[H]
\centering{}\caption{\linespread{0.50}\selectfont{}\label{fig: Angular simple implementation}Implementation
of angular averaging of two Gaussian CDFs. In (a), a diagonal straight
line is drawn (in black), and on it are marked many evenly spaced
points. Our empirical work used 1001 such points. In (b), an angled
straight line at angle $\theta=65^{\textrm{o}}$ is drawn (in grey)
through each marked point and between the two CDFs. In (c), the midpoint
of each angled straight line is obtained, and these midpoints map
out the new CDF, with linear interpolation used to produce the full
continuous CDF.}
\vspace{-0.2cm}
\includegraphics[scale=0.47]{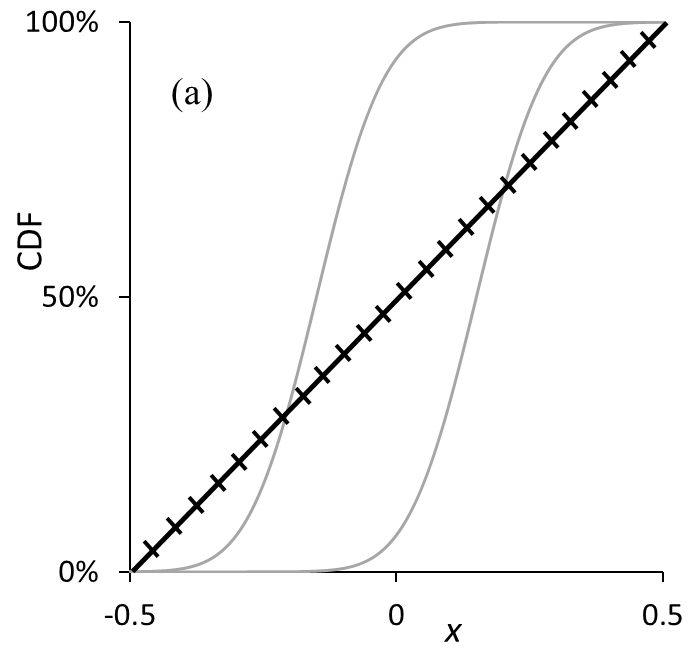}\includegraphics[scale=0.47]{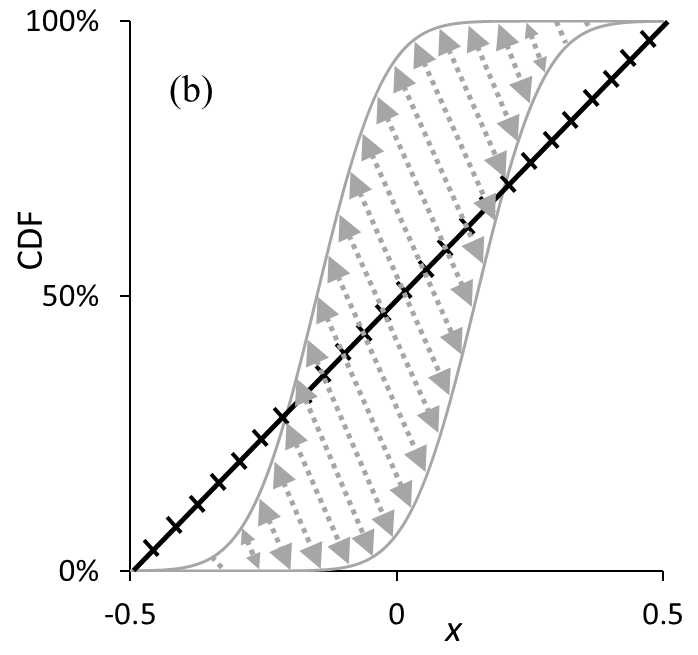}\includegraphics[scale=0.47]{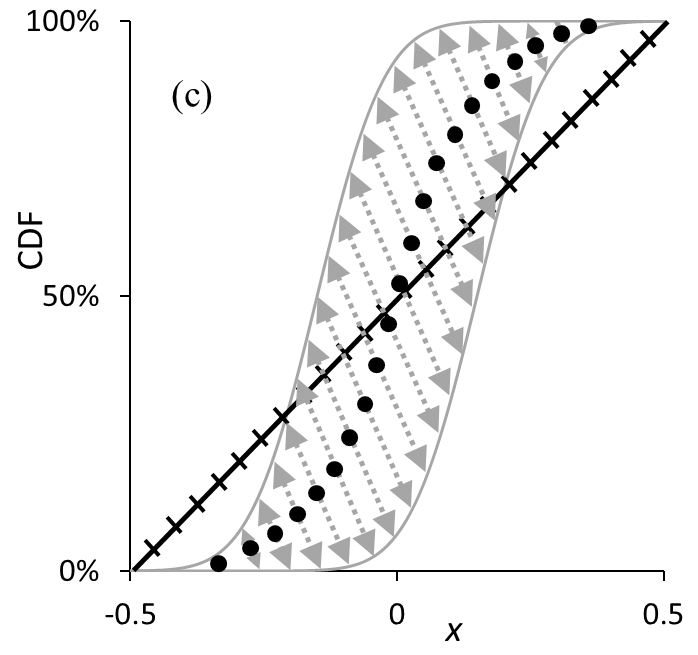}\vspace{-0.8cm}
\end{figure}

To implement angular combining, we propose two numerical approaches.
The first is a simple pragmatic approach based on the geometrical
interpretation in Figure~\ref{fig: Angular averaging basics}. In
essence, it involves identifying the intersection points of the angled
line $y=-\tan\theta(x-c)$ with each CDF $F_{i}$, and then computing
the weighted average of the points of intersection for each value
of $c$. Note that the intersection of each angled line with each
CDF typically lacks a closed-form solution, thus necessitating a numerical
method to determine the intersection point. To convey how we implemented
this approach in our empirical analysis, Figure~\ref{fig: Angular simple implementation}
provides an illustration for angular averaging at angle $\theta=65^{\textrm{o}}$
using the same two Gaussian CDFs averaged in our previous figures.
Notice that we have scaled the $x$-axes so that they each have unit
length. This enables us to compare more easily values of the angle
used for applications to different data. For CDFs with unbounded support,
we would pragmatically suggest the scaled $x$-axis of unit length
extends to at least the 1\% and 99\% quantiles of each CDF.

The second approach\textbf{ }to implementation is a simulation algorithm
for generating randomly sampled values for the distribution produced
by angular combining. An appeal of this approach is that it avoids
the need to solve implicit equations to find the points of intersection
of the angled lines with the CDFs, which is required in the pragmatic
approach of Figure~\ref{fig: Angular simple implementation}. The
simulation algorithm is \textcolor{blue}{described }in Online Appendix
C\textcolor{blue}{, and available in an R package at: https://github.com/XiaochunMeng1/R-package-for-Angular-Combining}.\medskip{}

\section{Theoretical Results\label{Section 4}}

In this section, we present results concerning the mean, variance,
intervals and CRPS of the distributional forecast resulting from angular
combining. We also consider the distribution produced by an angular
form of median aggregation. All proofs are provided in Online Appendix
A. 

\subsection{Mean of Angular Combining\label{Section 4.1}}

\citet{lichtendahl2013better} show that, for vertical and horizontal
averaging, the means of their resulting CDFs both equal the average
of the means of the individual CDFs. They explain that this is an
attractive property because the mean of a CDF forecast is widely used
as a point forecast, and averaging is often a successful approach
to aggregating point forecasts. The property is not possessed by many
other methods for combining CDF forecasts, such as median aggregation
and the logarithmic opinion pool. It is, therefore, interesting that
angular averaging possesses this property. This is implied by Proposition~\ref{Prop: Mean},
which considers the general case of weighted combining.

\begin{proposition}\label{Prop: Mean}For horizontal, vertical and
angular combining, with any set of weights $\{w_{1},...w_{k}\}$,
the mean of the combination is identical to the weighted average of
the means of the individual CDFs $F_{i}$ in the combination.

\end{proposition}

In empirical studies, CDF forecast accuracy is often heavily influenced
by the location of the CDF forecasts. A practical implication of the
combining methods delivering CDFs with the same mean is that their
relative accuracy will be due to other aspects of the CDF, such as
the variance.

\subsection{Variance of Angular Combining\label{Section 4.2}}

\citet{lichtendahl2013better} prove that the horizontal average \textcolor{blue}{has
lower variance} than the vertical average. Theorem~\ref{Thm 1: Var ang =000026 vert}
extends this result to angular combining, of which angular averaging
is a special case. 

\begin{theorem}\label{Thm 1: Var ang =000026 vert}For any set of
weights $\left\{ w_{1},...w_{k}\right\} $, the angular combination
$F_{A,w,\theta}$ \textcolor{blue}{has lower variance} than the vertical
combination $F_{V,w}$.

\end{theorem}

Theorem~\ref{Thm 1: Var ang =000026 vert} implies that the angular
and horizontal combinations both \textcolor{blue}{have lower variance}
than the vertical combination. In fact, it is straightforward to extend
the proof of this theorem regarding variance to higher even (central)
moments. As angular combining conceptually lies between vertical and
horizontal combining, it is natural to ask whether the variance of
the angular combination falls between the variances of the vertical
and horizontal combinations. While this may not hold generally, we
now explain that it is true in the specific scenario described in
Assumption~\ref{Assumption}.

\begin{assumption}\label{Assumption}Each individual distribution
$F_{i}$ in the combination has the same scale and is from the same
location-scale family, where the PDF of the standardized member of
the location-scale family is symmetric about 0 and unimodal.

\end{assumption}

These assumptions are among those adopted in Proposition 5 of \citet{lichtendahl2013better},
which compares the CDFs for vertical and horizontal averaging. Our
Lemma~\ref{Lem: Shape} below extends the proposition to the angular
averaging of two CDFs.

\begin{lemma}\label{Lem: Shape}Under Assumption~\ref{Assumption},
when two distributions are averaged, the PDF $f_{A,\theta}$ of angular
averaging is symmetric about its mean, and $F_{A,\theta}(x)$ is increasing
w.r.t. $\theta$ when $x$ is below the mean, and $F_{A,\theta}(x)$
is decreasing w.r.t. $\theta$ when $x$ is above the mean.

\end{lemma}

Theorem~\ref{Thm 2: Var ang decreasing} builds on Lemma~\ref{Lem: Shape}
to show that the variance of the angular combination $F_{A,w,\theta}$
is increasing w.r.t. the angle $\theta$, regardless of the number
of CDFs included in the combination.

\begin{theorem}\label{Thm 2: Var ang decreasing}Under Assumption~\ref{Assumption},
for any set of weights $\{w_{1},...,w_{k}\}$, the variance of the
angular combination $F_{A,w,\theta}$ is increasing w.r.t. $\theta$,
with the implication that the horizontal combination \textcolor{blue}{has
lower variance} than the angular combination, which \textcolor{blue}{has
lower variance} than the vertical combination.

\end{theorem}

Proposition~\ref{Prop: Mean} and Theorem~\ref{Thm 2: Var ang decreasing}
lead to practical guidance on selecting the best combining method
w.r.t. the mean and variance, under Assumption~\ref{Assumption}.
Using historical data, we first set the mean by obtaining the weights
$w_{i}$, and then perturb $\theta$ to target the variance. The weights
can be optimized using a point forecast combining method, or, as in
our empirical study, pragmatically set to be inversely proportional
to the CRPS. Once the weights are determined, we produce the CDF forecasts
from vertical and horizontal combining for each historical period.
If the variances of both are too high for the past data, horizontal
combining should be used; if the variances of both are too low, vertical
combining is recommended; and if the variances of the CDFs produced
by horizontal combining are too low, while the variances of the CDFs
from vertical combining are too high, there exists a value of $\theta$
for which angular combining delivers the CDF forecasts with most accurate
variances.

\subsection{Interval Forecasts from Angular Averaging\label{Section 4.3}}

In this section, we consider the interval forecasts, containing the
medians, obtained from the CDFs produced by the different combining
approaches. \citet{lichtendahl2013better} show that under Assumption~\ref{Assumption},
if the means $\mu_{i}$ of individual predictions are symmetrically
located around the common mean of the horizontal and vertical averages,
horizontal averaging results in a narrower interval compared to vertical
combining. Proposition~\ref{Prop: Coverage 1} extends this, showing
that the interval forecasts of horizontal averaging are also narrower
than those for angular averaging.

\begin{proposition}\label{Prop: Coverage 1}Let $\alpha_{1}<0.5$
and $\alpha_{2}>0.5$ denote two arbitrary probability values. Under
Assumption~\ref{Assumption}, if the $\mu_{i}$ are symmetrically
located around the common mean of the horizontal and angular averages,
we have $\left(F_{H}^{-1}(\alpha_{1}),F_{H}^{-1}(\alpha_{2})\right)\subseteq\left(F_{A,\theta}^{-1}(\alpha_{1}),F_{A,\theta}^{-1}(\alpha_{2})\right)$.

\end{proposition}

Proposition~\ref{Prop: Coverage 2}, as a direct consequence of Lemma~\ref{Lem: Shape},
provides a stronger result when two CDFs are averaged. In this case,
vertical and horizontal averaging produce the widest and narrowest
prediction intervals, respectively, while angular averaging yields
intervals that fall between the two,

\begin{proposition}\label{Prop: Coverage 2}Let $\alpha_{1}<0.5$
and $\alpha_{2}>0.5$ denote two arbitrary probability values. Under
Assumption~\ref{Assumption}, the interval forecast for the angular
average of two CDFs expands as the angle $\theta$ increases, i.e.,
$\left(F_{A,\theta_{1}}^{-1}(\alpha_{1}),F_{A,\theta_{1}}^{-1}(\alpha_{2})\right)\subseteq\left(F_{A,\theta_{2}}^{-1}(\alpha_{1}),F_{A,\theta_{2}}^{-1}(\alpha_{2})\right)$
if $\theta_{1}\leq\theta_{2}$.

\end{proposition}

This proposition can be used as the basis for similar practical guidance
to that presented in Section~\ref{Section 4.2}, with the best choice
between horizontal, vertical and angular averaging dictated by the
interval forecasts derived from the CDF forecasts produced by horizontal
and vertical averaging. \textcolor{blue}{We note that,} in comparison
with the intervals of the data generating process (DGP), wider intervals
can be viewed as indicating underconfidence, while narrower intervals
imply overconfidence (see, for example, \citealt{Moore2008,lichtendahl2013better,Soll2023}). 

\subsection{Scoring Rules and Angular Combining\label{Section 4.4}}

Let $Z$ denote the random variable of the DGP, and let $G$ denote
its CDF. A scoring rule $S(F,z)$ assigns a numerical score to a CDF
forecast $F$ and a realization $z$, and is \textit{proper} if its
expectation is minimized when $F$ coincides with $G$. Hence, proper
scoring rules, such as the CRPS, incentivize honest reporting by forecasters.
The \textit{divergence} of a proper scoring rule is defined as $\mathbb{E}\left[S(F,Z)\right]-\mathbb{E}\left[S(G,Z)\right]$
(\citealt{gneiting2007strictly}). The divergence is a nonnegative
quantity that measures the `distance' between $F$ and $G$, with
a smaller value indicating superior accuracy. By contrast with variance
and interval forecasts, which are inherent properties of a CDF forecast,
the value of a score depends on the DGP, making it impossible to determine
the best model without knowledge of the DGP. Nevertheless, we can
still make the meaningful observation that, as angular combining encompasses
both horizontal and vertical combining, given sufficient data to estimate
$\theta$, angular combining will not be outperformed by horizontal
or vertical combining for any proper score, i.e., these combining
methods have greater `distance' from the true DGP than angular combining.

\citet{lichtendahl2013better} show that the score for vertical averaging
is no greater than the average of the scores of the individual distributions
for the linear score, log score, quadratic score and CRPS. They show
that this is also true for the linear and log scores when it comes
to horizontal averaging. Although, for the quadratic score, the result
is not necessarily valid for horizontal averaging, the authors establish
conditions under which it is true. They do not discuss the result
for horizontal averaging for the CRPS. Theorem~\ref{Thm: CRPS angular}
addresses this, and extends the result to angular combining. Before
providing the theorem, we define the CRPS, with $\I(\cdot)$ representing
the indicator function,
\begin{align*}
CRPS(F,z) & =\intop_{-\infty}^{\infty}\left(F(x)-\I(x>z)\right)^{2}dx.
\end{align*}
\vspace{-0.9cm}

\begin{theorem}\label{Thm: CRPS angular}For any set of weights $\{w_{1},...,w_{k}\}$,
the CRPS for the horizontal and angular combinations are less than
or equal to the weighted combination of the CRPS of the individual
distributional forecasts $F_{i}$ in the combination.

\end{theorem}

This theorem indicates that angular averaging captures the wisdom
of the crowd by providing ``accuracy that is no worse than the average
member of the crowd'' (\citealt{Larrick2012}; \citealt{lichtendahl2013better}).
A further practical implication of Theorem~\ref{Thm: CRPS angular}
is that angular combining cannot be outperformed, in terms of the
CRPS, by the weakest individual CDF forecast. The theorem also has
practical relevance for numerically optimizing the combining weights
based on the CRPS, because it ensures that any local minimum is also
the global optimum. Consequently, one can initialize the numerical
procedure with a set of arbitrary weights and employ a local optimization
technique to obtain the optimal weights.

\subsection{Angular Median Aggregation\label{Section 4.5}}

\citet{Hora2013} show that performing median aggregation horizontally
and vertically results in the same CDF. They find that, although for
a set of calibrated individual CDF forecasts, median aggregation will
deliver a CDF that is not calibrated and is overdispersed, the miscalibration
is less than with vertical averaging. The approach has the appeal
of robustness and computational simplicity, which motivates consideration
of an angular form of median aggregation. 

Similarly to angular averaging, the median can be obtained at any
angle $\theta$. In other words, to construct the angular median CDF
for a chosen $\theta$, we find the median of the points of intersection
of the individual CDFs with an angled line $y=-\tan\theta(x-c)$,
and repeat this for different values of $c\in(-\infty,\infty)$. Theorem~\ref{Thm: Median}
shows that, if there is an odd number of individual CDFs, the same
CDF is produced when the median is obtained vertically, horizontally
or at an angle.

\begin{theorem}\label{Thm: Median}When $k$ is odd, obtaining the
median of $k$ CDFs, horizontally, vertically or at an angle $\theta$
produces the same CDF.

\end{theorem}\medskip{}

\section{Empirical Study of Covid Mortality Forecasting\label{Section 5}}

In this section, we use mortality data to evaluate our proposed approach
to distributional forecast combining. 

\subsection{The Dataset\label{Section 5.1}}

We used data from the COVID-19 Forecast Hub (https://covid19forecasthub.org/).
The Hub provides open access to weekly observations and forecasts
from multiple forecasting teams for one to four weeks-ahead for the
number of reported Covid deaths at the national and state level in
the U.S. Mortality forecasts were used by public health decision makers
during the pandemic (\citealt{Adam2020}; \citealt{Ray2023}). Forecasts
from the Hub gained widespread attention with weekly updates presented
on the website of the Center for Disease Control and Prevention. The
teams submitting forecasts to the Hub came from academia, industry
and government affiliated groups. We used forecasts of the number
of weekly Covid deaths produced from the 84 weekly forecast origins
from 6 June 2020 to 8 January 2022, inclusive, and mortality observations,
recorded on 15 January 2022, for the weeks ending 13 June 2020 to
15 January 2022, inclusive. We considered mortality at the national
level, for the 50 states, and for the District of Columbia, which
for simplicity we refer to as a state. The dataset, therefore, involved
52 time series. \citet{Taylor2022} used the same data in a study
of interval forecast combining. 

\begin{figure}[H]
\centering{}\caption{\linespread{0.50}\selectfont{}\label{fig: Box plots for number of teams}For
each of the 84 forecast origins in the Covid dataset, the box plot
summarizes the number of eligible forecasting teams for each of the
52 series.}
\vspace{-0.2cm}
\includegraphics[scale=0.4]{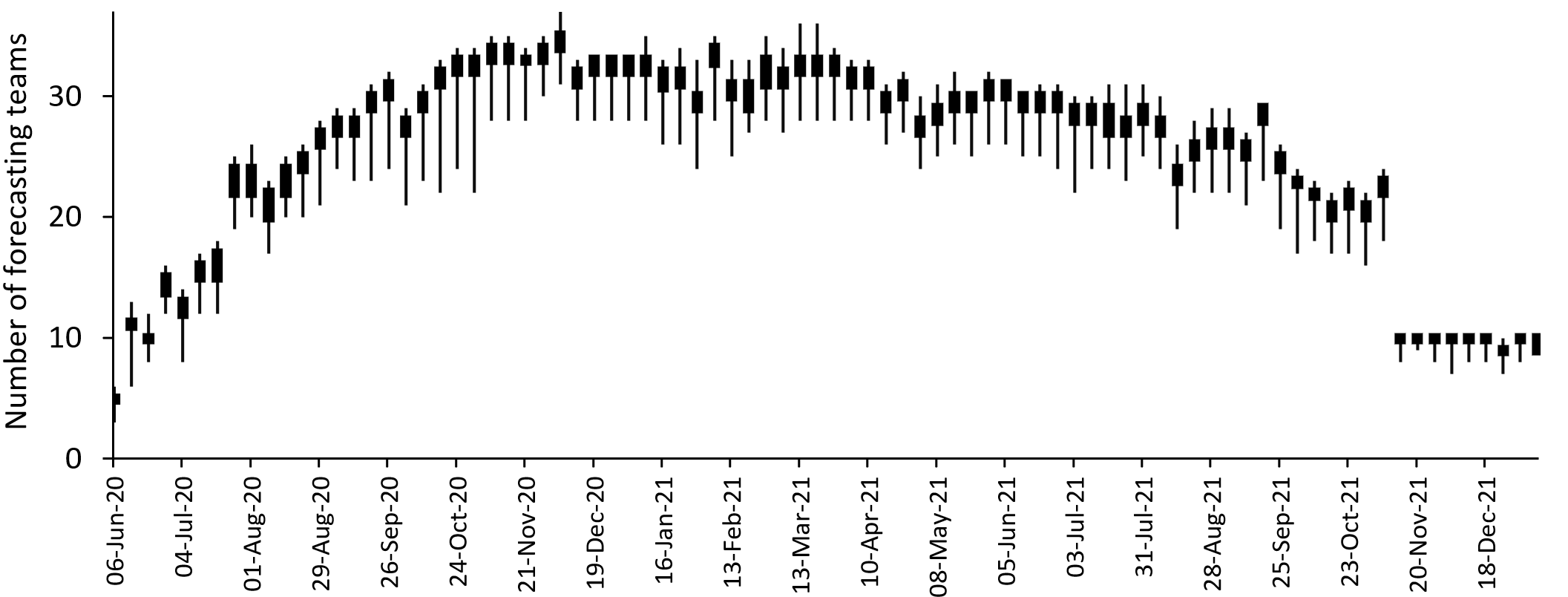}\vspace{-0.8cm}
\end{figure}

The curators of the Hub invited forecasting teams to submit a point
forecast and quantile forecasts corresponding to the following 23
probability levels: 1\%, 2.5\%, 5\%, 10\%, 15\%, 20\%, 25\%, 30\%,
35\%, 40\%, 45\%, 50\%, 55\%, 60\%, 65\%, 70\%, 75\%, 80\%, 85\%,
90\%, 95\%, 97.5\% and 99\%. As our interest in this paper is in distributional
forecasts, we constructed continuous CDFs from the 23 quantiles. To
do this, we used the following simple approach: we defined the lower
bound of the distribution as being a distance below the 1\% quantile
equal to the difference between the 1\% and 2.5\% quantiles (unless
this resulted in a negative lower bound, in which case we set it to
be zero); we defined the upper bound of the distribution as being
a distance above the 99\% quantile equal to the difference between
the 97.5\% and 99\% quantiles; and we used linear interpolation to
give the CDF between each pair of quantiles, between the lower bound
and the 1\% quantile, and between the 99\% quantile and the upper
bound. (Defining the lower and upper bounds as the 1\% and 99\% quantiles,
respectively, produced very similar out-of-sample results for the
evaluation measures we describe in Section~\ref{Section 5.2}.) Each
week, the curators of the Hub generated an \textquotedblleft ensemble\textquotedblright{}
forecast combination. Initially, the ensemble was the horizontal average;
from mid-July 2020, it was computed as the median; and after mid-November
2021, it was a horizontal weighted combination. At each forecast origin,
for each series, the curators of the Hub screened forecasting teams
for inclusion in their ensemble, removing teams if their forecasts
were clearly unrealistic. We followed the same approach, so that,
for each forecast origin and series, we included in combinations only
the teams considered eligible by the Hub curators for that forecast
origin and series. There was variation over time in the identity and
number of teams. In Figure~\ref{fig: Box plots for number of teams},
for each forecast origin, a box plot summarizes the differing number
of eligible teams for each of the 52 series. About half the teams
used compartmental models. The others employed statistical methods,
machine learning or agent-based models. The diversity of methods motivates
attempts to extract wisdom from the crowd (see \citealt{Surowiecki2004}).

\medskip{}

\subsection{Structure of the Study and Evaluation Measures\label{Section 5.2}}

We used forecasts made from the first 10 out of the 84 forecast origins
for the initial in-sample parameter estimation. For the remaining
74 forecast origins, we used an expanding in-sample period consisting
of all weeks up to and including the origin itself. Out-of-sample
forecasts were, therefore, produced for the final 74 weeks of each
of the 52 mortality time series. \textcolor{blue}{Our choice of just
10 periods for the initial estimation sample was driven by the desire
for a reasonably long out-of-sample period.}

We used the following multiple quantile score (MQS) to assess distributional
forecast accuracy,
\[
MQS\left(F,z\right)=\frac{1}{23}\sum_{i=1}^{23}2QS\left(F^{-1}(\alpha_{i}),z\right)\quad\textrm{where}\quad QS\left(F^{-1}(\alpha_{i}),z\right)=\left(\alpha_{i}-\I(z\leq F^{-1}(\alpha_{i}))\right)\left(z-F^{-1}(\alpha_{i})\right).
\]
$z$ is the observation, $F$ is the CDF, and $F^{-1}(\alpha_{i})$
is the quantile with probability level $\alpha_{i}$, which we set
as the 23 probability levels chosen by the Hub curators. $QS$ is
the widely-used score for a quantile, which is also known as the \textit{pinball}
or \textit{check} loss function. The MQS is a proper score for the
distribution (\citealt{grushka2017quantile}; \citealt{jose2009evaluating}).
It approximates the CRPS. We use the MQS rather than the CRPS itself,
as the MQS is employed in other studies of the COVID-19 Forecast Hub
data, which reflects the focus of the Hub on forecasting the 23 quantiles
(see, for example, \citealt{Bracher2021}; \citealt{Ray2023}). In
addition to the MQS, we also evaluated the QS for each of the 23 probability
levels to provide insight into forecast accuracy for the different
regions of the distributions.

In addition to averaging each score over the out-of-sample period,
we also averaged over the lead times, as the relative performances
of the methods were similar for the four lead times, and the series
were relatively short. We report the scores averaged across all 52
series, but as this will be heavily influenced by the higher mortality
series, we also report results for the following four groupings of
the series: the U.S. national level series; the 17 states for which
cumulative mortality was highest in the final week of the dataset;
the 17 states with the next highest cumulative mortality in the final
week; and the 17 states with lowest cumulative mortality in the final
week. 

We also computed skill scores, defined as the percentage by which
a method is more accurate than a benchmark, which we chose to be vertical
averaging. Skill scores provide a further way of avoiding higher mortality
series dominating when averaging across series. To average over skill
scores, we first calculated the geometric mean of the ratios of the
score for each method to the score for the benchmark method, then
subtracted this from 1, and multiplied the result by 100.

To evaluate unconditional calibration for the CDF forecasts, we used
reliability diagrams, which convey calibration across the quantiles
of the CDF. 

\subsection{Implementing Horizontal, Vertical and Angular Combining\label{Section 5.3}}

We implemented horizontal, vertical and angular averaging, as well
as horizontal, vertical and angular forms of performance-based weighted
combining. As we discussed in relation to Figure~\ref{fig: Box plots for number of teams},
the identity and number of teams submitting forecasts to the Hub varied
over time. In view of this, we pragmatically set each combining weight
to be inversely proportional to the team\textquoteright s in-sample
MQS, which was one of the combining approaches considered in the horizontal
combining study of \citet{Taylor2023}. If fewer than five past periods
were available from which to produce the in-sample MQS for a team,
the MQS was assumed equal to the average of the MQS of the other teams
in the combination. Note that the weights only approximately reflect
relative performance because the MQS will typically not have been
computed from the same past periods for all teams, due to inconsistencies
in the record of submissions among the teams.

Our implementation of horizontal averaging and horizontal weighted
combining simply involved combining the quantile forecasts for each
of the 23 probability levels. Vertical combining involved averaging
and weighted combining of the probability forecasts for each of a
set of values of the mortality variable. This set consisted of all
23 quantile forecasts submitted by all the teams.

To implement angular combining, we used the pragmatic numerical approach
described in Section~\ref{Section 3.4}. This was adequate for our
study because identifying the intersection points of each angled line
with each CDF forecast was straightforward due to the CDF forecasts
being piecewise-linear. In Figure~\ref{fig: Angular simple implementation}
of Section~\ref{Section 3.4}, we described how, for angular averaging,
the approach involved, for each of the $m$=1001 angled straight lines
(in grey), averaging the points of intersection of the angled straight
line with the CDF forecasts. For each angled straight line, in addition
to averaging, we applied the performance-based weighted combination,
using the same weights we had used for horizontal and vertical weighted
combining. We then optimized the angle $\theta$ separately for angular
averaging and angular weighted combining. For each series and method,
at each forecast origin, we optimized $\theta$ by minimizing the
in-sample MQS averaged over the four lead times, with the in-sample
MQS for each lead time computed using only observations at or before
the forecast origin. 

\begin{figure}[H]
\centering{}\caption{\linespread{0.50}\selectfont{}\label{fig: Angular CDF for California =000026 New York}1
week-ahead CDF forecasts for Covid mortality in the week up to 15
January 2022 from 10 individual teams and the three averaging methods
for (a) California, and (b) New York State. For angular averaging,
the optimized angle $\theta$ was $60^{\textrm{o}}$ for California
and $79^{\textrm{o}}$ for New York.}
\vspace{-0.2cm}
\includegraphics[scale=0.75]{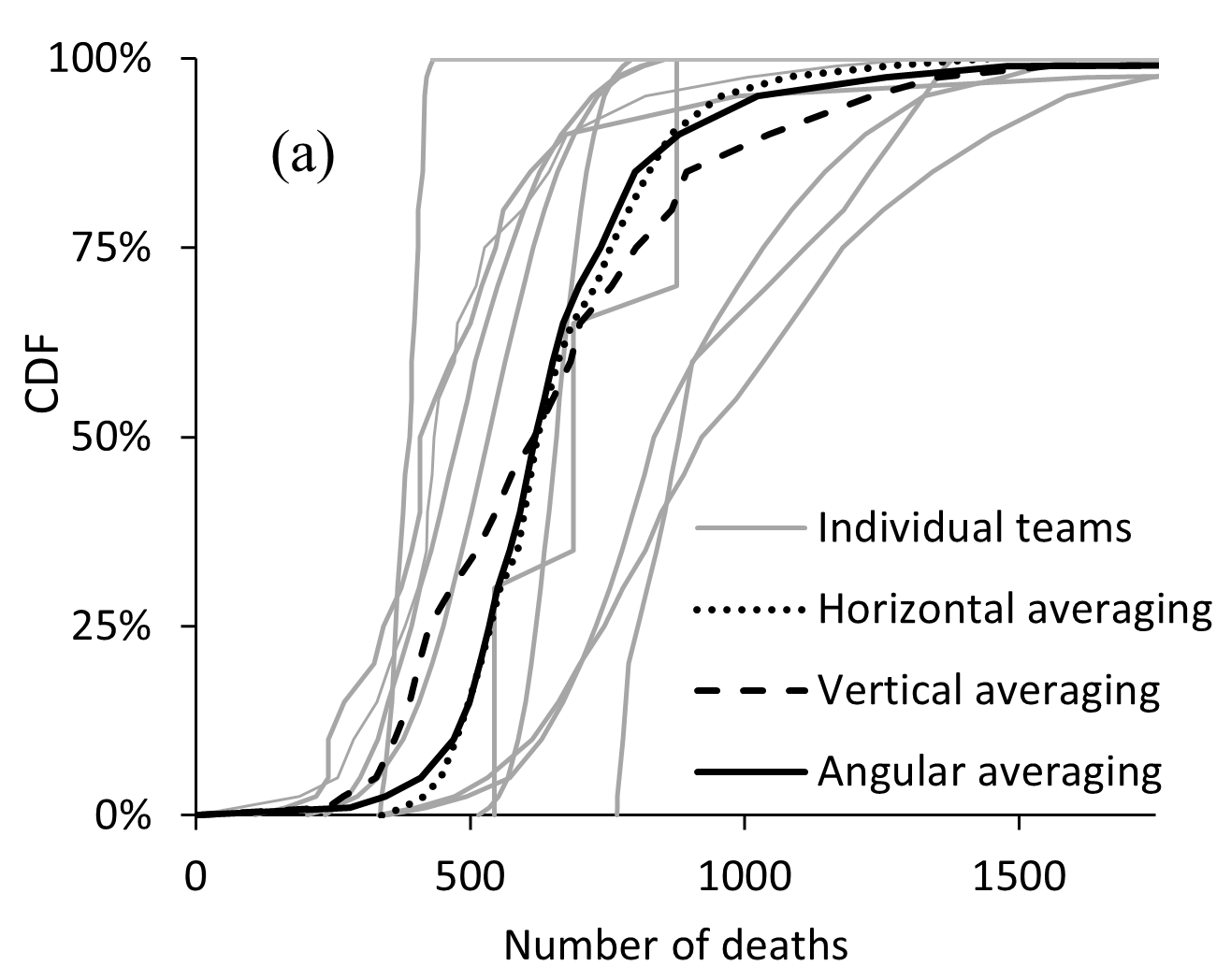}\includegraphics[scale=0.75]{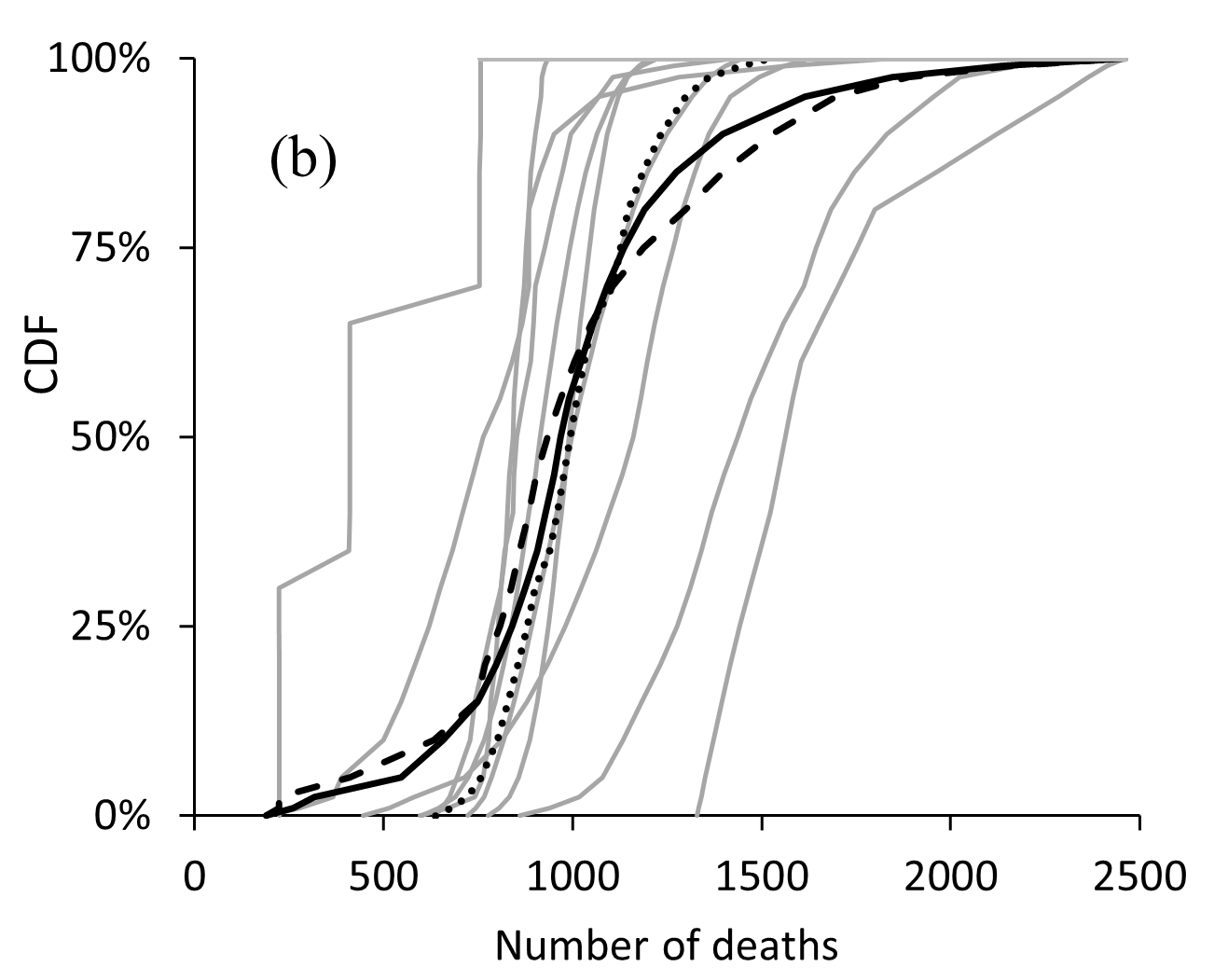}\vspace{-0.8cm}
\end{figure}

Figure~\ref{fig: Angular CDF for California =000026 New York} provides
examples of the CDF forecasts produced by the individual teams and
by horizontal, vertical and angular averaging. The examples correspond
to the forecasts produced from the final forecast origin in our dataset
for California and New York State. For each state, the figure shows
that the CDF forecasts of the teams vary quite considerably in terms
of their location, spread and shape. We can also see differences between
the CDFs produced by the three forms of averaging.

In a small number of cases, forecasting teams submitted CDF forecasts
for which the quantiles were identical for two or more adjacent values
of the 23 probability levels. For example, this is the case for the
left-most CDF forecast in Figure~\ref{fig: Angular CDF for California =000026 New York}(b).
This is problematic for vertical combining because, for certain values
on the $x$-axis, there is more than \textcolor{blue}{one} CDF \textcolor{blue}{value}.
To address this, for any such value on the $x$-axis, we simply averaged
the CDF values before applying vertical combining.

With optimization of the angle $\theta$ performed at each forecast
origin, angular combining essentially allows switching between different
angles as the method is passed through the out-of-sample period. To
challenge this, we implemented a method that switched between horizontal
and vertical combining, with the choice between these two combinations
made, at each forecast origin, by selecting the one with lowest in-sample
MQS. We call this \textit{horizontal/vertical switching}. We implemented
a version of this that switched between horizontal and vertical averaging,
and, separately, a version that switched between horizontal and vertical
weighted combining.

As our focus in this paper is horizontal, vertical and angular combining,
we initially report results for only these methods. In Section~\ref{Section 5.8},
we present results for other combining methods. 

Comparing the results of the combining methods with those of individual
teams is difficult because none of the teams provided forecasts that
passed the Hub\textquoteright s eligibility screening for all past
periods and series. In view of this, we focus only on combining methods
in our analysis.

\subsection{Results for Angular Combining with a Fixed Angle\label{Section 5.4}}

\begin{figure}[H]
\centering{}\caption{\linespread{0.50}\selectfont{}\label{fig: MQS for different angles}For
the Covid dataset, MQS skill score for the out-of-sample period for
angular averaging at different angles in (a), and for angular weighted
combining in (b). Higher values of the skill score are better.}
\vspace{-0.2cm}
\includegraphics[scale=0.72]{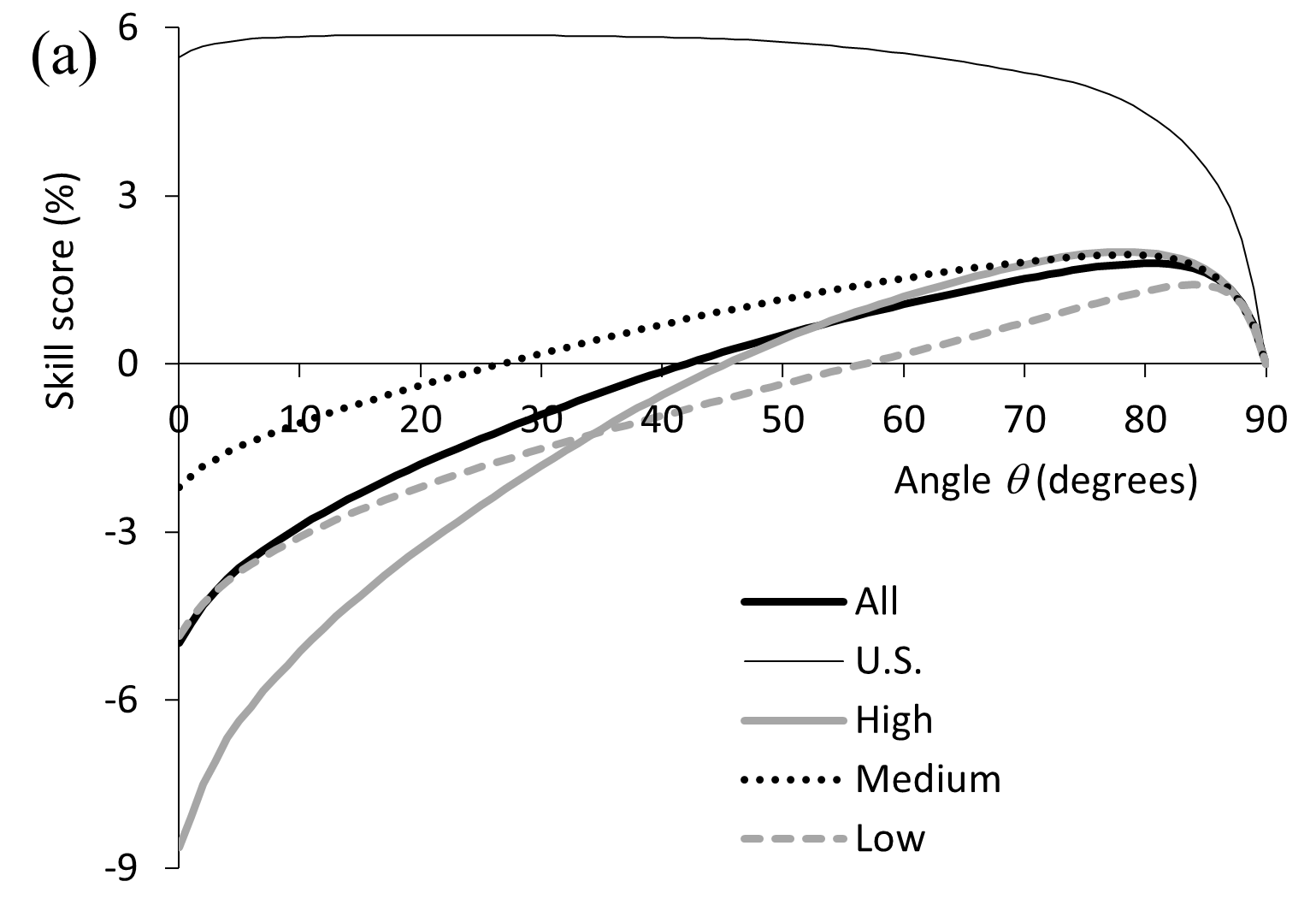}\includegraphics[scale=0.72]{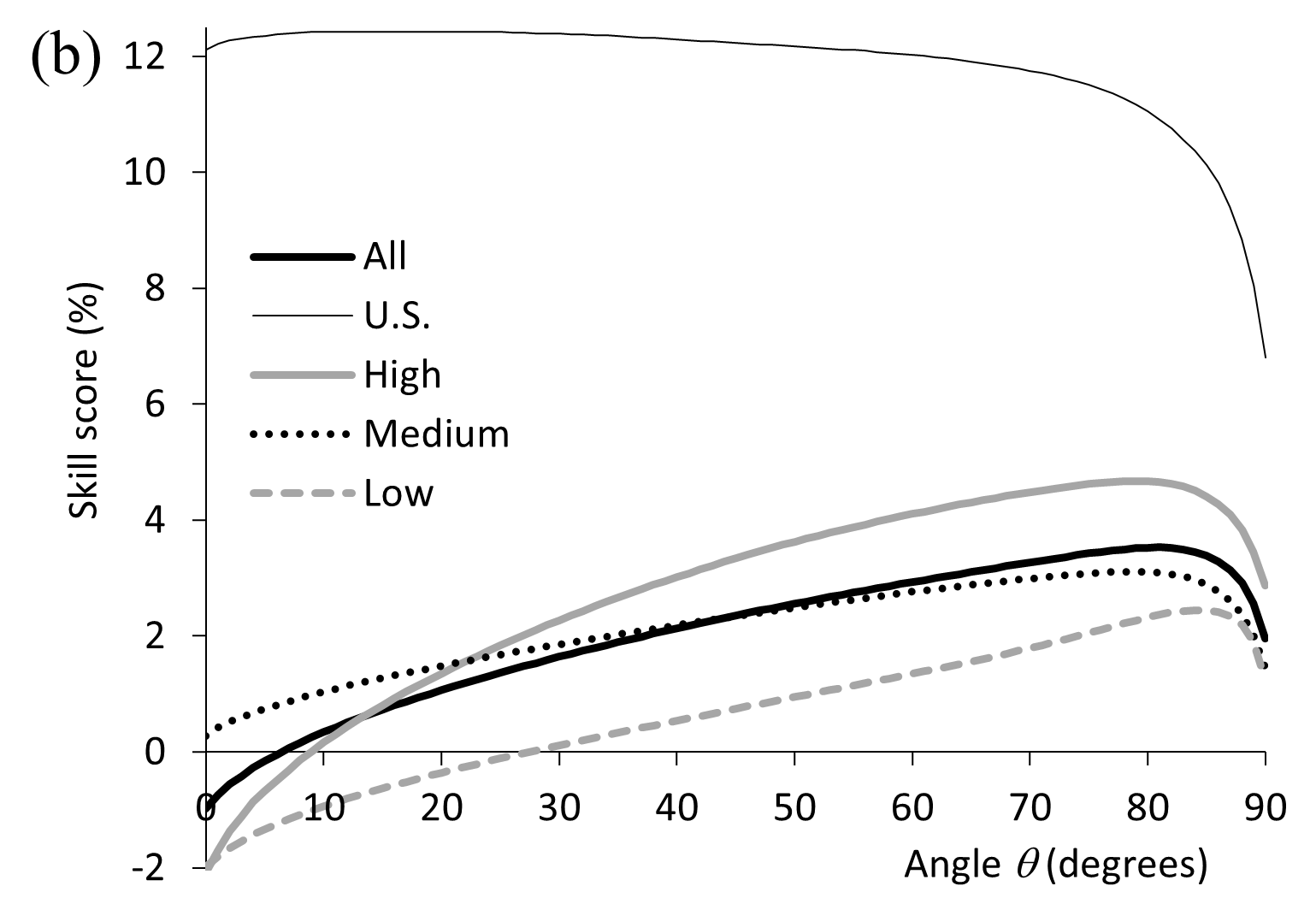}\vspace{-0.7cm}
\end{figure}

As an initial investigation into suitable values of the angle $\theta$
used in angular combining, we evaluated the accuracy of the approach
when $\theta$ was not optimized but was fixed at the same value for
all 74 forecast origins used to produce out-of-sample forecasts. We
did this, in turn, for integer values of $\theta$ between $0^{\textrm{o}}$
and $90^{\textrm{o}}$. For angular averaging, Figure~\ref{fig: MQS for different angles}(a)
presents the resulting MQS skill scores, computed for the five categories
of series: all 52 series, the U.S. national level series, and the
high, medium and low mortality groupings of series. Higher values
of the skill score are preferable. First note that the extreme right
of the figure ($\theta=90^{\textrm{o}}$) corresponds to vertical
averaging, and as this is the benchmark method in the skill score
calculation, the skill scores are all zero at this right extreme.
The left extreme ($\theta=0^{\textrm{o}}$) corresponds to horizontal
averaging. For the U.S. national level series, the figure shows the
skill score relatively high for values of $\theta$ up to about $70^{\textrm{o}}$,
beyond which the skill score reduces to its lowest value at $\theta=90^{\textrm{o}}$,
indicating vertical averaging is not suitable for this mortality series.
For the other four categories of series, the other four lines in Figure~\ref{fig: MQS for different angles}(a)
show the skill score at its lowest when $\theta=0^{\textrm{o}}$,
indicating that angular averaging with any angle above $0^{\textrm{o}}$
was more accurate than horizontal averaging. These four lines all
peak with $\theta$ between about $80^{\textrm{o}}$ and $85^{\textrm{o}}$,
suggesting that angular averaging with an angle in this range would
be suitable for series in these four categories, and that it would
be more accurate than vertical averaging ($\theta=90^{\textrm{o}}$).
 Figure~\ref{fig: MQS for different angles}(b) provides similar
insight for angular weighted combining to that provided by Figure~\ref{fig: MQS for different angles}(a).

The message from Figure~\ref{fig: MQS for different angles} that
relatively high values for $\theta$ tended to be most suitable is
interesting because, for the angular averaging of two Gaussian CDFs
in Figure~\ref{fig: Angular averaging CDF =000026 PDF for different angles},
an informal conclusion could be that the CDF resulting from angular
averaging only became very noticeably different to horizontal averaging
for values of $\theta$ above about $70^{\textrm{o}}$. However, such
a conclusion from Figure~\ref{fig: Angular averaging CDF =000026 PDF for different angles}
ignores the tails of the distributions, which are noticeably different
for much lower values of $\theta$. It should also not be overlooked
that Figure~\ref{fig: MQS for different angles} presents the MQS,
which summarizes accuracy for the full distribution. This motivates
a more careful consideration of the tails of the distribution produced
by angular averaging, and we do this using the quantile score in \textcolor{blue}{Section}~\ref{Section 5.7}.

\subsection{Optimized Values of the Angle\label{Section 5.5}}

\begin{figure}[H]
\centering{}\caption{\linespread{0.50}\selectfont{}\label{fig: Histogram for angles}For
angular averaging and angular weighted combining, histogram of the
angle $\theta$ optimized for the 52 Covid series at the 74 forecast
origins.}
\vspace{-0.2cm}
\includegraphics[scale=0.8]{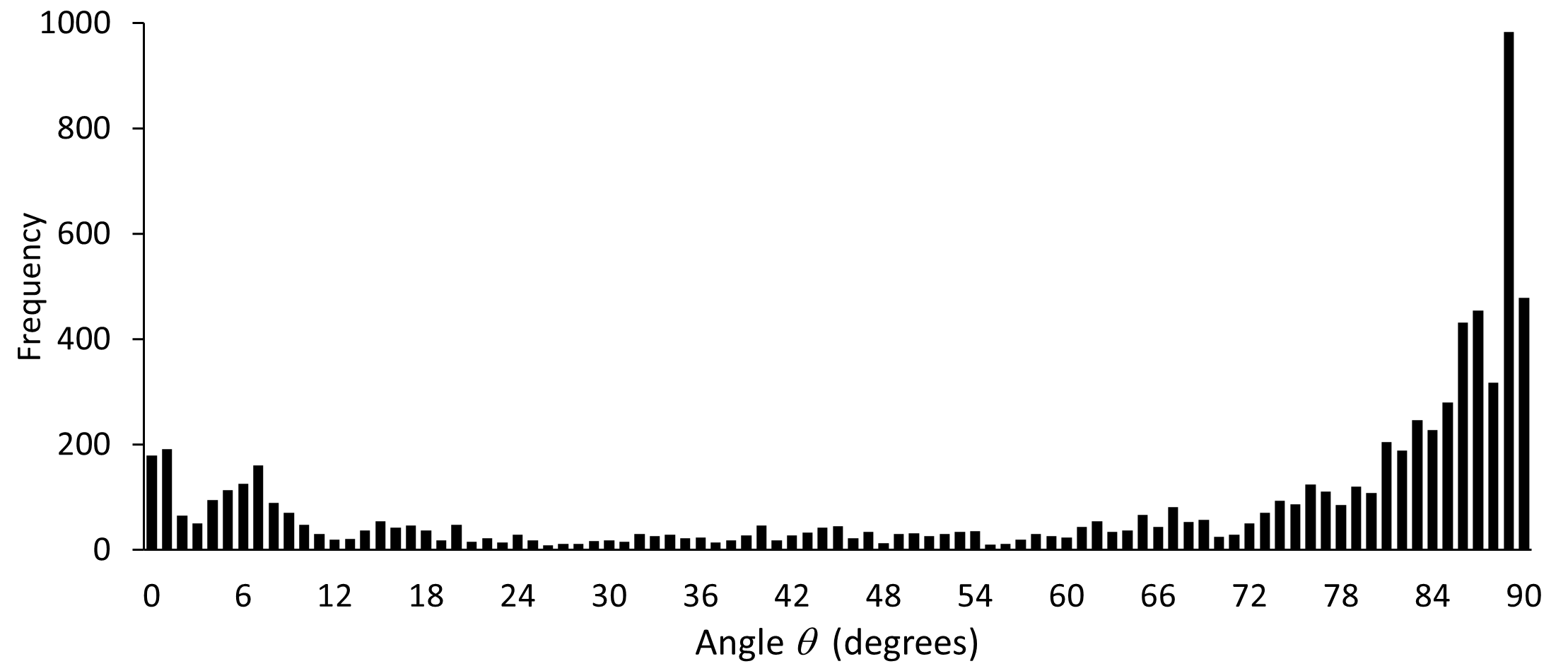}\vspace{-0.8cm}
\end{figure}

In practice, the angle $\theta$ would need to be estimated. In our
empirical comparison of methods in Sections~\ref{Section 5.6} to
\ref{Section 5.8}, we implemented angular averaging and angular weighted
combining for each of the 52 series and 74 forecast origins used to
produce out-of-sample forecasts. For each of these methods, we optimized
the angle $\theta$ for each series and origin by minimizing the in-sample
MQS. The histogram in Figure~\ref{fig: Histogram for angles} summarizes
the resulting optimized values of $\theta$. The left extreme of the
figure shows that angles close to the horizontal ($\theta=0^{\textrm{o}}$)
were quite often found to be optimal. However, the right end of the
figure shows that the optimized angle was more frequently found to
be closer to vertical ($\theta=90^{\textrm{o}}$), with angles between
about $70^{\textrm{o}}$ and $90^{\textrm{o}}$ often obtained.

\begin{figure}[H]
\centering{}\caption{\linespread{0.50}\selectfont{}\label{fig: Angles time series}For
angular averaging, the optimized angle $\theta$ at the 74 forecast
origins for five of the Covid series.}
\vspace{-0.2cm}
\includegraphics[scale=0.75]{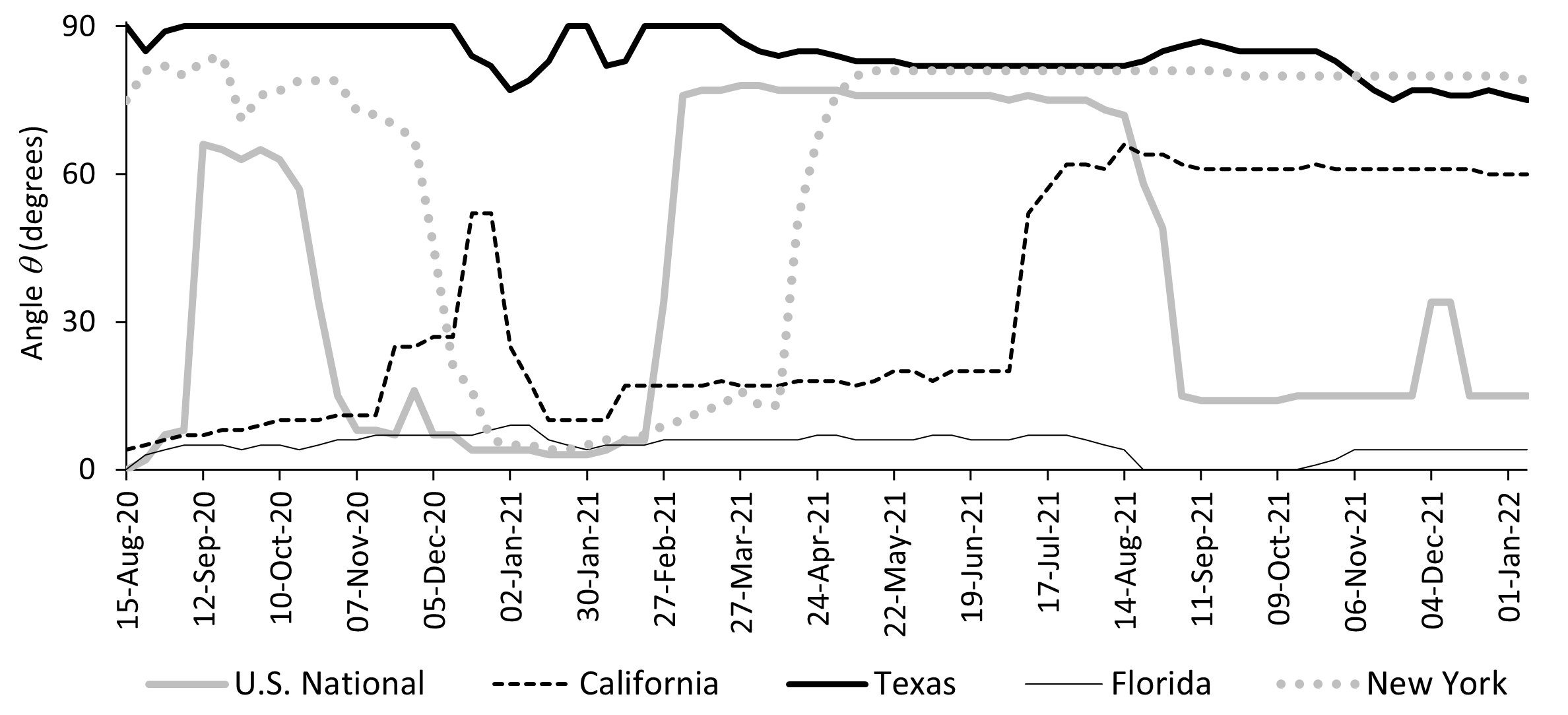}\vspace{-0.6cm}
\end{figure}

Figure~\ref{fig: Angles time series} shows the angles optimized
for each forecast origin for the five series for which mortality was
highest in the final period of the dataset. The figure shows that
for Florida and Texas, the optimized angles remained quite close to
$0^{\textrm{o}}$ and $90^{\textrm{o}}$, respectively, across the
forecast origins. By contrast the optimized angles for the other three
series were generally not particularly close to the extremes of $0^{\textrm{o}}$
and $90^{\textrm{o}}$. 

\subsection{Comparing Horizontal, Vertical and Angular Combining using the MQS\label{Section 5.6}}

In this section, we evaluate the out-of-sample performance of angular
combining with optimized angle by comparing it with horizontal and
vertical combining, as well as horizontal/vertical switching. Table~\ref{Table 1: MQS}
summarizes the MQS results. The first five columns of values provide
the MQS for each of the five categories of the time series considered
in Figure~\ref{fig: MQS for different angles}. Lower values of the
MQS are better. The final five columns of Table~\ref{Table 1: MQS}
present the corresponding skill scores. In each column, bold highlights
the best approach to averaging and the best approach to weighted combining.
Looking at the averaging approaches in the first four rows, we see
that horizontal averaging was the most accurate for the U.S. national
level series and the poorest for the other four categories. The horizontal/vertical
switching method performed well for the U.S. national level series
because it selected horizontal averaging for all forecast origins.
For the other four categories, the skill scores show that the switching
method was not able to outperform vertical averaging. For these four
categories, angular averaging was a little more accurate than vertical
averaging. For weighted combining, we again see horizontal combining
performing well only for the U.S. national level series, with angular
combining again a little better than vertical combining overall. We
also note that each form of weighted combining outperforms the corresponding
form of averaging. In Online Appendix D.1, for the methods in Table~\ref{Table 1: MQS},
we present the skill scores for each of the 52 series. 

\begin{table}[H]
{\footnotesize\caption{\linespread{0.50}\selectfont{}\label{Table 1: MQS}For the Covid
dataset, MQS for the out-of-sample period for the three forms of combining
and horizontal/vertical switching.}
\vspace{0.1cm}
}\linespread{0.50}\selectfont{}%
\begin{tabular}{lccccccccccc}
\toprule 
 &  &  & {\footnotesize\textsf{MQS}} &  &  & ~~~ & \multicolumn{5}{c}{{\footnotesize\textsf{MQS skill score (\%)}}}\tabularnewline
 & {\footnotesize\textsf{U.S.}} & {\footnotesize\textsf{High}} & {\footnotesize\textsf{Medium}} & {\footnotesize\textsf{Low}} & {\footnotesize\textsf{All}} &  & {\footnotesize\textsf{U.S.}} & {\footnotesize\textsf{High}} & {\footnotesize\textsf{Medium}} & {\footnotesize\textsf{Low}} & {\footnotesize\textsf{All}}\tabularnewline
\midrule
{\footnotesize\textsf{\textit{Averaging}}} &  &  &  &  &  &  &  &  &  &  & \tabularnewline
{\footnotesize\textsf{~~Vertical}} & {\footnotesize\textsf{949.1}} & {\footnotesize\textsf{69.2}} & {\footnotesize\textsf{27.7}} & {\footnotesize\textsf{\textbf{8.2}}} & {\footnotesize\textsf{52.6}} &  & {\footnotesize\textsf{0.0}} & {\footnotesize\textsf{0.0}} & {\footnotesize\textsf{0.0}} & {\footnotesize\textsf{0.0}} & {\footnotesize\textsf{0.0}}\tabularnewline
{\footnotesize\textsf{~~Horizontal}} & {\footnotesize\textsf{\textbf{897.3}}} & {\footnotesize\textsf{80.1}} & {\footnotesize\textsf{28.5}} & {\footnotesize\textsf{8.5}} & {\footnotesize\textsf{55.5}} &  & {\footnotesize\textsf{\textbf{5.5}}} & {\footnotesize\textsf{-8.6}} & {\footnotesize\textsf{-2.2}} & {\footnotesize\textsf{-4.9}} & {\footnotesize\textsf{-5.0}}\tabularnewline
{\footnotesize\textsf{~~Horizontal/vertical switching}} & {\footnotesize\textsf{\textbf{897.3}}} & {\footnotesize\textsf{69.8}} & {\footnotesize\textsf{27.9}} & {\footnotesize\textsf{\textbf{8.2}}} & {\footnotesize\textsf{51.9}} &  & {\footnotesize\textsf{\textbf{5.5}}} & {\footnotesize\textsf{-1.3}} & {\footnotesize\textsf{-0.7}} & {\footnotesize\textsf{-1.3}} & {\footnotesize\textsf{-1.0}}\tabularnewline
{\footnotesize\textsf{~~Angular}} & {\footnotesize\textsf{903.3}} & {\footnotesize\textsf{\textbf{68.5}}} & {\footnotesize\textsf{\textbf{27.5}}} & {\footnotesize\textsf{\textbf{8.2}}} & {\footnotesize\textsf{\textbf{51.5}}} &  & {\footnotesize\textsf{4.8}} & {\footnotesize\textsf{\textbf{1.0}}} & {\footnotesize\textsf{\textbf{0.9}}} & {\footnotesize\textsf{\textbf{0.1}}} & {\footnotesize\textsf{\textbf{0.8}}}\tabularnewline
\midrule 
{\footnotesize\textsf{\textit{Weighted combining}}} &  &  &  &  &  &  &  &  &  &  & \tabularnewline
{\footnotesize\textsf{~~Vertical}} & {\footnotesize\textsf{884.5}} & {\footnotesize\textsf{67.1}} & {\footnotesize\textsf{27.3}} & {\footnotesize\textsf{\textbf{8.1}}} & {\footnotesize\textsf{50.5}} &  & {\footnotesize\textsf{6.8}} & {\footnotesize\textsf{2.9}} & {\footnotesize\textsf{1.4}} & {\footnotesize\textsf{\textbf{1.2}}} & {\footnotesize\textsf{2.0}}\tabularnewline
{\footnotesize\textsf{~~Horizontal}} & {\footnotesize\textsf{\textbf{834.1}}} & {\footnotesize\textsf{75.2}} & {\footnotesize\textsf{28.0}} & {\footnotesize\textsf{8.2}} & {\footnotesize\textsf{52.5}} &  & {\footnotesize\textsf{\textbf{12.1}}} & {\footnotesize\textsf{-2.1}} & {\footnotesize\textsf{0.3}} & {\footnotesize\textsf{-2.0}} & {\footnotesize\textsf{-1.0}}\tabularnewline
{\footnotesize\textsf{~~Horizontal/vertical switching}} & {\footnotesize\textsf{\textbf{834.1}}} & {\footnotesize\textsf{67.3}} & {\footnotesize\textsf{27.6}} & {\footnotesize\textsf{\textbf{8.1}}} & {\footnotesize\textsf{49.7}} &  & {\footnotesize\textsf{\textbf{12.1}}} & {\footnotesize\textsf{3.1}} & {\footnotesize\textsf{1.0}} & {\footnotesize\textsf{0.0}} & {\footnotesize\textsf{1.6}}\tabularnewline
{\footnotesize\textsf{~~Angular}} & {\footnotesize\textsf{839.1}} & {\footnotesize\textsf{\textbf{66.6}}} & {\footnotesize\textsf{\textbf{27.2}}} & {\footnotesize\textsf{\textbf{8.1}}} & {\footnotesize\textsf{\textbf{49.4}}} &  & {\footnotesize\textsf{11.6}} & {\footnotesize\textsf{\textbf{4.0}}} & {\footnotesize\textsf{\textbf{2.4}}} & {\footnotesize\textsf{1.1}} & {\footnotesize\textsf{\textbf{2.7}}}\tabularnewline
\bottomrule
\end{tabular}

{\small\vspace{-0.6cm}
}{\small\par}
\end{table}
{\scriptsize\textsf{Note: The unit of the score is deaths. Lower }}{\scriptsize\textsf{\textcolor{blue}{scores}}}{\scriptsize\textsf{
and higher skill }}{\scriptsize\textsf{\textcolor{blue}{scores}}}{\scriptsize\textsf{
are better. Skill }}{\scriptsize\textsf{\textcolor{blue}{scores use}}}{\scriptsize\textsf{
vertical averaging as benchmark. Bold indicates the best method in
each column within the averaging block and within the weighted combining
block.}}{\scriptsize\par}

In Section~\ref{Section 4.1}, we noted that the CDFs produced by
horizontal, vertical and angular combining have the same mean if the
same combining weights are used. This is useful because it implies
that their relative performances in Table~\ref{Table 1: MQS} are
due to other aspects of the CDFs, such as their \textcolor{blue}{variance}.

In some applications, \textcolor{blue}{data is not available} to optimize
combining method parameters. From Table~\ref{Table 1: MQS}, vertical
or horizontal averaging could be used, although the choice between
the two would need to be made subjectively. For angular averaging,
when the angle $\theta$ cannot be optimized, a simple idea, informally
lying midway between horizontal and vertical averaging, is to select
$\theta=45^{\textrm{o}}$. \textcolor{blue}{This would seem reasonable
if the forecaster had no basis on which to choose between horizontal
and vertical averaging. However, if they had prior experience with
similar datasets, this could lead to }a moderate preference for either
horizontal or vertical averaging, \textcolor{blue}{in which case}
they could, perhaps, select $\theta=22.5^{\textrm{o}}$ or $\theta=67.5^{\textrm{o}}$,
respectively. In view of Theorems~\ref{Thm 1: Var ang =000026 vert}
and \ref{Thm 2: Var ang decreasing}, another approach would be to
choose the angle $\theta$ that delivers a CDF with the desired standard
deviation. For example, $\theta$ could be chosen so that $\sigma_{A}$
is the average of $\sigma_{H}$ and $\sigma_{V}$, where $\sigma_{A}$,
$\sigma_{H}$ and $\sigma_{V}$ are the standard deviations of the
CDFs produced by angular, horizontal and vertical averaging, respectively.
If the forecaster has a view as to whether $\sigma_{H}$ or $\sigma_{V}$
is closer to the truth, then $\theta$ could, perhaps, be chosen so
that $\sigma_{A}=0.75\sigma_{H}+0.25\sigma_{V}$ or $\sigma_{A}=0.25\sigma_{H}+0.75\sigma_{V}$,
respectively. Table~\ref{Table 2: MQS for data-free methods} compares
the accuracy of these proposals. The table shows that angular averaging
with $\theta=45^{\textrm{o}}$ performed reasonably, but the best
performing method was angular averaging with $\theta=67.5^{\textrm{o}}$,
confirming insight from Figure~\ref{fig: MQS for different angles}
that, overall, angular averaging was, for this dataset, most accurate
for values of the angle $\theta$ above about $45^{\textrm{o}}$.

\begin{table}[H]
{\footnotesize\caption{\linespread{0.50}\selectfont{}\label{Table 2: MQS for data-free methods}For
the Covid dataset, MQS for the out-of-sample period for combining
methods involving no parameter optimization.}
\vspace{0.1cm}
}\linespread{0.50}\selectfont{}%
\begin{tabular}{lccccccccccc}
\toprule 
 & \multicolumn{5}{c}{{\footnotesize\textsf{MQS}}} & ~~~ & \multicolumn{5}{c}{{\footnotesize\textsf{MQS skill score (\%)}}}\tabularnewline
 & {\footnotesize\textsf{U.S.}} & {\footnotesize\textsf{High}} & {\footnotesize\textsf{Medium}} & {\footnotesize\textsf{Low}} & {\footnotesize\textsf{All}} &  & {\footnotesize\textsf{U.S.}} & {\footnotesize\textsf{High}} & {\footnotesize\textsf{Medium}} & {\footnotesize\textsf{Low}} & {\footnotesize\textsf{All}}\tabularnewline
\midrule
\multicolumn{12}{l}{{\footnotesize\textsf{\textit{Averaging (from Table \ref{Table 1: MQS})}}}}\tabularnewline
{\footnotesize\textsf{~~Vertical}} & {\footnotesize\textsf{949.1}} & {\footnotesize\textsf{\textbf{69.2}}} & {\footnotesize\textsf{27.7}} & {\footnotesize\textsf{8.2}} & {\footnotesize\textsf{52.6}} &  & {\footnotesize\textsf{0.0}} & {\footnotesize\textsf{0.0}} & {\footnotesize\textsf{0.0}} & {\footnotesize\textsf{0.0}} & {\footnotesize\textsf{0.0}}\tabularnewline
{\footnotesize\textsf{~~Horizontal}} & {\footnotesize\textsf{897.3}} & {\footnotesize\textsf{80.1}} & {\footnotesize\textsf{28.5}} & {\footnotesize\textsf{8.5}} & {\footnotesize\textsf{55.5}} &  & {\footnotesize\textsf{5.5}} & {\footnotesize\textsf{-8.6}} & {\footnotesize\textsf{-2.2}} & {\footnotesize\textsf{-4.9}} & {\footnotesize\textsf{-5.0}}\tabularnewline
\midrule
{\footnotesize\textsf{\textit{Angular averaging}}} &  &  &  &  &  &  &  &  &  &  & \tabularnewline
{\footnotesize\textsf{~~Angle $\theta=22.5^{\textrm{o}}$}} & {\footnotesize\textsf{\textbf{893.5}}} & {\footnotesize\textsf{75.0}} & {\footnotesize\textsf{28.0}} & {\footnotesize\textsf{8.3}} & {\footnotesize\textsf{53.6}} &  & {\footnotesize\textsf{\textbf{5.9}}} & {\footnotesize\textsf{-2.9}} & {\footnotesize\textsf{-0.2}} & {\footnotesize\textsf{-2.0}} & {\footnotesize\textsf{-1.6}}\tabularnewline
{\footnotesize\textsf{~~Angle $\theta=45^{\textrm{o}}$}} & {\footnotesize\textsf{894.1}} & {\footnotesize\textsf{72.4}} & {\footnotesize\textsf{27.7}} & {\footnotesize\textsf{8.2}} & {\footnotesize\textsf{52.6}} &  & {\footnotesize\textsf{5.8}} & {\footnotesize\textsf{0.0}} & {\footnotesize\textsf{0.9}} & {\footnotesize\textsf{-0.6}} & {\footnotesize\textsf{0.2}}\tabularnewline
{\footnotesize\textsf{~~Angle $\theta=67.5^{\textrm{o}}$}} & {\footnotesize\textsf{898.9}} & {\footnotesize\textsf{70.2}} & {\footnotesize\textsf{27.5}} & {\footnotesize\textsf{\textbf{8.1}}} & {\footnotesize\textsf{\textbf{51.9}}} &  & {\footnotesize\textsf{5.3}} & {\footnotesize\textsf{\textbf{1.6}}} & {\footnotesize\textsf{\textbf{1.8}}} & {\footnotesize\textsf{\textbf{0.6}}} & {\footnotesize\textsf{\textbf{1.4}}}\tabularnewline
{\footnotesize\textsf{~~$\sigma_{A}=0.75\sigma_{H}+0.25\sigma_{V}$}} & {\footnotesize\textsf{898.2}} & {\footnotesize\textsf{74.6}} & {\footnotesize\textsf{27.9}} & {\footnotesize\textsf{8.3}} & {\footnotesize\textsf{53.5}} &  & {\footnotesize\textsf{5.4}} & {\footnotesize\textsf{-3.3}} & {\footnotesize\textsf{0.2}} & {\footnotesize\textsf{-2.0}} & {\footnotesize\textsf{-1.6}}\tabularnewline
{\footnotesize\textsf{~~$\sigma_{A}=0.5\sigma_{H}+0.5\sigma_{V}$}} & {\footnotesize\textsf{905.3}} & {\footnotesize\textsf{73.0}} & {\footnotesize\textsf{27.6}} & {\footnotesize\textsf{8.2}} & {\footnotesize\textsf{53.0}} &  & {\footnotesize\textsf{4.6}} & {\footnotesize\textsf{-1.7}} & {\footnotesize\textsf{1.0}} & {\footnotesize\textsf{-0.9}} & {\footnotesize\textsf{-0.4}}\tabularnewline
{\footnotesize\textsf{~~$\sigma_{A}=0.25\sigma_{H}+0.75\sigma_{V}$}} & {\footnotesize\textsf{919.2}} & {\footnotesize\textsf{71.2}} & {\footnotesize\textsf{\textbf{27.4}}} & {\footnotesize\textsf{8.2}} & {\footnotesize\textsf{52.6}} &  & {\footnotesize\textsf{3.2}} & {\footnotesize\textsf{-0.2}} & {\footnotesize\textsf{1.5}} & {\footnotesize\textsf{0.0}} & {\footnotesize\textsf{0.5}}\tabularnewline
\bottomrule
\end{tabular}

{\small\vspace{-0.6cm}
}{\small\par}
\end{table}

{\scriptsize\textsf{Note: The unit of the score is deaths. Lower }}{\scriptsize\textsf{\textcolor{blue}{scores}}}{\scriptsize\textsf{
and higher skill }}{\scriptsize\textsf{\textcolor{blue}{scores}}}{\scriptsize\textsf{
are better. Skill }}{\scriptsize\textsf{\textcolor{blue}{scores use}}}{\scriptsize\textsf{
vertical averaging from Table~\ref{Table 1: MQS} as benchmark. Bold
indicates the best method in each column. }}{\scriptsize\par}

\begin{figure}[H]
\centering{}\caption{\linespread{0.50}\selectfont{}\label{fig: 23QuScores}Quantile skill
scores averaged across all 52 Covid series for the 23 individual quantiles
for the out-of-sample period for the averaging methods in (a), and
weighted combining methods in (b). Higher values of the skill score
are better.}
\vspace{-0.2cm}
\includegraphics[scale=0.72]{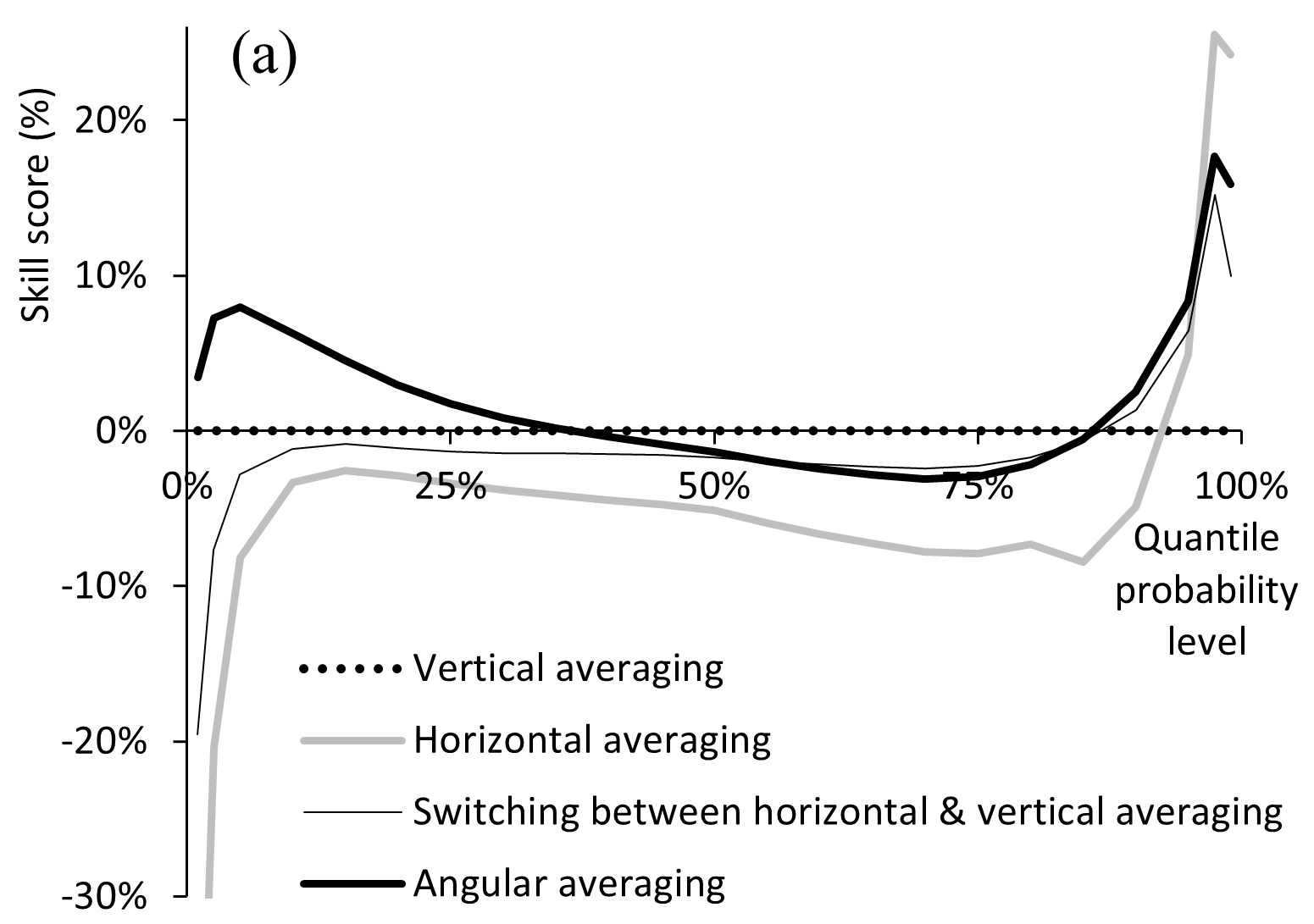}\includegraphics[scale=0.72]{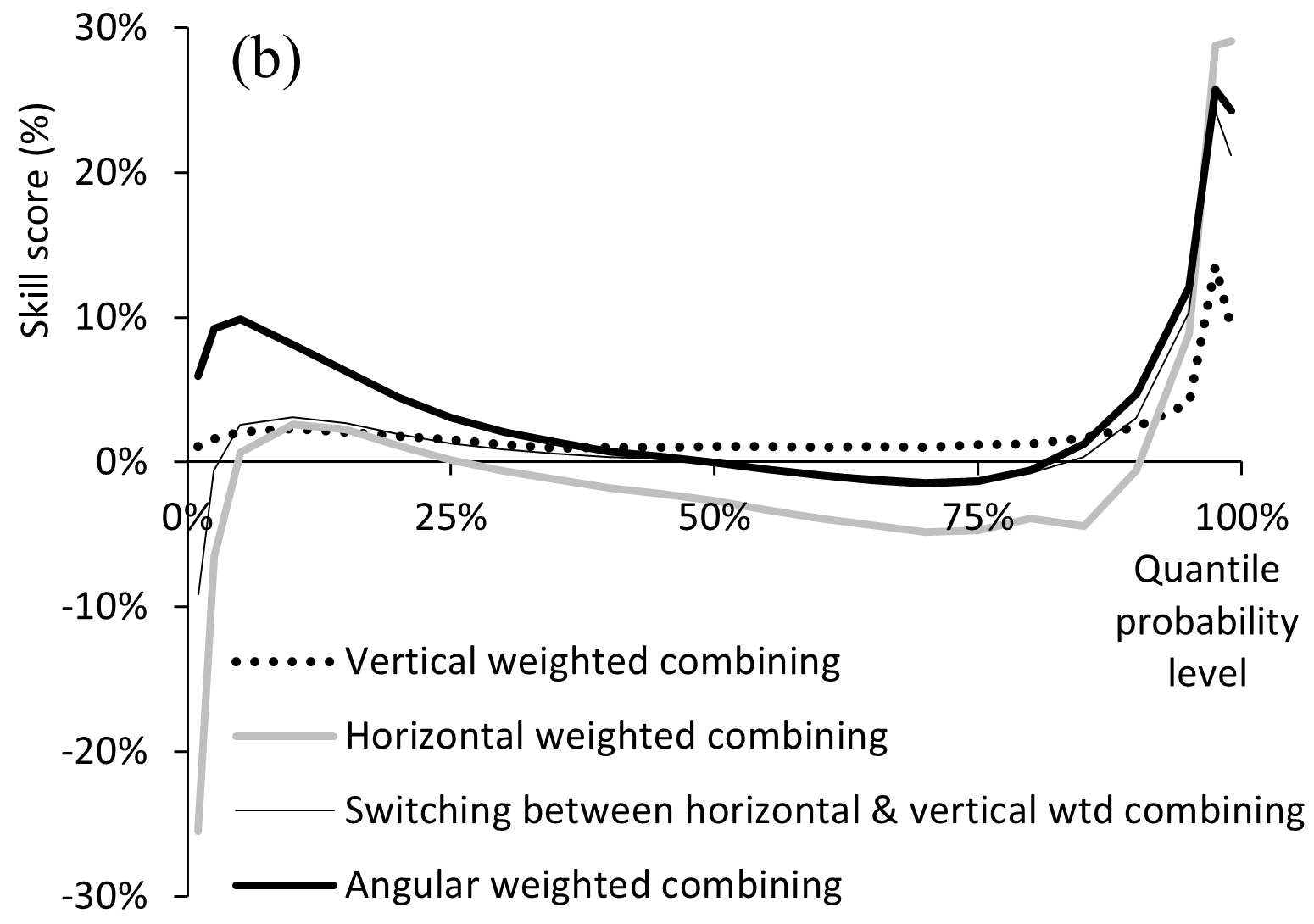}\vspace{-0.4cm}
\end{figure}

\subsection{Comparing Horizontal, Vertical and Angular Combining for Different
Quantiles\label{Section 5.7}}

Figure~\ref{fig: 23QuScores} reports quantile skill scores for the
23 probability levels for the methods in Table~\ref{Table 1: MQS}.
The skill scores have been computed using the vertical average as
the benchmark, and averaged across the out-of-sample period for all
52 series. In view of the results in Table~\ref{Table 1: MQS} for
horizontal averaging, it is not surprising to see that negative skill
scores dominate Figure~\ref{fig: 23QuScores}(a) for this method.
However, the positive skill scores for this method for the extreme
upper quantiles indicate that, for at least part of the distributions,
horizontal was more accurate than vertical averaging. It is interesting
to compare the results for vertical and angular combining in Figures~\ref{fig: 23QuScores}(a)
and (b). Both plots show particularly strong results for angular combining
in the tails of the distributions, while vertical combining was more
competitive for the central quantiles. We can infer from this that
for interval forecasts with a high nominal coverage probability, such
as 95\%, angular combining would be preferable for this dataset, while
vertical and angular combining would both be reasonable choices for
50\% intervals. In Online Appendix D.2, we provide plots showing the
quantile skill scores separately for the U.S. national level series,
and the high, medium and low mortality groupings of series.

\textcolor{blue}{We also evaluated accuracy across the 23 quantiles
of each CDF forecast using unconditional calibration. We did this
using reliability diagrams, which we present in Online Appendix D.3.} 

\subsection{Evaluating Other Methods using the MQS\label{Section 5.8}}

Although our aim in this paper is to introduce angular combining as
an alternative to horizontal and vertical combining, it is also interesting
to see how other methods from the literature perform for the mortality
dataset. Table~\ref{Table 3: MQS for various methods} reports the
results for the methods listed below, along with the results from
Table 1 for angular averaging and angular weighted combining, which
we have positioned at the top of Table~\ref{Table 3: MQS for various methods}.
Method parameters were optimized using the same approach that we employed
to optimize $\theta$ for angular combining. For each series and forecast
origin, this involved minimizing the in-sample MQS averaged over the
four lead times. 

\textit{Median aggregation} -- This is the simple method considered
by \citet{Hora2013}, which, as we showed in Section~\ref{Section 4.5},
produces the same CDF if implemented horizontally, vertically or at
an angle.

\textit{Beta-transformed linear pool} -- \citet{gneiting2013combining}
use the beta distribution to transform the CDF forecast of a vertical
weighted combination. We \textcolor{blue}{implemented} this with weights
defined as in the weighted combinations of Table~\ref{Table 1: MQS}.
The implementation involved optimizing the two parameters of the beta
distribution. 

\textit{Trimming} -- We implemented the trimming approach of \citet{Jose2014}
that bases trimming on the mean of each CDF forecast. Exterior trimming
involves averaging the CDF forecasts that remain after removing a
percentage of the individual CDF forecasts with lowest and highest
mean. Interior trimming involves averaging the CDF forecasts that
were among a percentage of those with lowest and highest mean. We
implemented these methods with vertical, horizontal and angular averaging.
Each form of trimming required the optimization of the trimming percentage. 

\textit{Recalibration} -- \citet{Han2022} recalibrate a CDF forecast
by transforming the probabilities corresponding to predefined bins.
We applied the method to the three weighted combining methods in Table~\ref{Table 1: MQS}.
We defined the bins as the 24 intervals bounded by the quantiles corresponding
to the 23 probability levels of focus in the COVID-19 Forecast Hub,
and the lower and upper bounds of each CDF. The method involved optimizing
one parameter.

\begin{table}[H]
{\footnotesize\caption{\linespread{0.50}\selectfont{}\label{Table 3: MQS for various methods}For
the Covid dataset, MQS for the out-of-sample period for a variety
of combining and recalibration methods.}
\vspace{0.1cm}
}\linespread{0.50}\selectfont{}%
\begin{tabular}{lccccccccccc}
\toprule 
 & \multicolumn{5}{c}{{\footnotesize\textsf{MQS}}} & ~~~~~ & \multicolumn{5}{c}{{\footnotesize\textsf{MQS skill score (\%)}}}\tabularnewline
 & {\footnotesize\textsf{U.S.}} & {\footnotesize\textsf{High}} & {\footnotesize\textsf{Medium}} & {\footnotesize\textsf{Low}} & {\footnotesize\textsf{All}} &  & {\footnotesize\textsf{U.S.}} & {\footnotesize\textsf{High}} & {\footnotesize\textsf{Medium}} & {\footnotesize\textsf{Low}} & {\footnotesize\textsf{All}}\tabularnewline
\midrule
\multicolumn{3}{l}{{\footnotesize\textsf{\textit{Angular combining (from Table \ref{Table 1: MQS})}}}} &  &  &  &  &  &  &  &  & \tabularnewline
{\footnotesize\textsf{~~Angular averaging}} & {\footnotesize\textsf{903.3}} & {\footnotesize\textsf{68.5}} & {\footnotesize\textsf{27.5}} & {\footnotesize\textsf{8.2}} & {\footnotesize\textsf{51.5}} &  & {\footnotesize\textsf{4.8}} & {\footnotesize\textsf{1.0}} & {\footnotesize\textsf{0.9}} & {\footnotesize\textsf{0.1}} & {\footnotesize\textsf{0.8}}\tabularnewline
{\footnotesize\textsf{~~Angular weighted combining}} & {\footnotesize\textsf{839.1}} & {\footnotesize\textsf{\textbf{66.6}}} & {\footnotesize\textsf{\textbf{27.2}}} & {\footnotesize\textsf{\textbf{8.1}}} & {\footnotesize\textsf{\textbf{49.4}}} &  & {\footnotesize\textsf{11.6}} & {\footnotesize\textsf{4.0}} & {\footnotesize\textsf{\textbf{2.4}}} & {\footnotesize\textsf{1.1}} & {\footnotesize\textsf{2.7}}\tabularnewline
\midrule 
\multicolumn{3}{l}{{\footnotesize\textsf{\textit{Simple benchmark and transformed linear
pool}}}} &  &  &  &  &  &  &  &  & \tabularnewline
{\footnotesize\textsf{~~Median}} & {\footnotesize\textsf{914.1}} & {\footnotesize\textsf{68.1}} & {\footnotesize\textsf{27.6}} & {\footnotesize\textsf{\textbf{8.1}}} & {\footnotesize\textsf{51.5}} &  & {\footnotesize\textsf{3.7}} & {\footnotesize\textsf{2.8}} & {\footnotesize\textsf{1.1}} & {\footnotesize\textsf{1.8}} & {\footnotesize\textsf{1.9}}\tabularnewline
{\footnotesize\textsf{~~Beta-transformed linear pool}} & {\footnotesize\textsf{\textbf{810.3}}} & {\footnotesize\textsf{71.3}} & {\footnotesize\textsf{28.2}} & {\footnotesize\textsf{8.5}} & {\footnotesize\textsf{50.9}} &  & {\footnotesize\textsf{\textbf{14.6}}} & {\footnotesize\textsf{0.2}} & {\footnotesize\textsf{-2.0}} & {\footnotesize\textsf{-3.6}} & {\footnotesize\textsf{-1.4}}\tabularnewline
\midrule
{\footnotesize\textsf{\textit{Exterior trimming}}} &  &  &  &  &  &  &  &  &  &  & \tabularnewline
{\footnotesize\textsf{~~Vertical}} & {\footnotesize\textsf{896.4}} & {\footnotesize\textsf{67.0}} & {\footnotesize\textsf{27.3}} & {\footnotesize\textsf{\textbf{8.1}}} & {\footnotesize\textsf{50.7}} &  & {\footnotesize\textsf{5.6}} & {\footnotesize\textsf{\textbf{4.3}}} & {\footnotesize\textsf{1.8}} & {\footnotesize\textsf{\textbf{2.1}}} & {\footnotesize\textsf{\textbf{2.8}}}\tabularnewline
{\footnotesize\textsf{~~Horizontal}} & {\footnotesize\textsf{897.0}} & {\footnotesize\textsf{79.3}} & {\footnotesize\textsf{27.8}} & {\footnotesize\textsf{8.2}} & {\footnotesize\textsf{55.0}} &  & {\footnotesize\textsf{5.5}} & {\footnotesize\textsf{-4.8}} & {\footnotesize\textsf{0.4}} & {\footnotesize\textsf{0.5}} & {\footnotesize\textsf{-1.2}}\tabularnewline
{\footnotesize\textsf{~~Angular}} & {\footnotesize\textsf{901.0}} & {\footnotesize\textsf{68.6}} & {\footnotesize\textsf{27.6}} & {\footnotesize\textsf{\textbf{8.1}}} & {\footnotesize\textsf{51.4}} &  & {\footnotesize\textsf{5.1}} & {\footnotesize\textsf{2.0}} & {\footnotesize\textsf{1.0}} & {\footnotesize\textsf{1.2}} & {\footnotesize\textsf{1.5}}\tabularnewline
\midrule
{\footnotesize\textsf{\textit{Interior trimming}}} &  &  &  &  &  &  &  &  &  &  & \tabularnewline
{\footnotesize\textsf{~~Vertical}} & {\footnotesize\textsf{951.0}} & {\footnotesize\textsf{72.3}} & {\footnotesize\textsf{27.7}} & {\footnotesize\textsf{8.2}} & {\footnotesize\textsf{53.7}} &  & {\footnotesize\textsf{-0.2}} & {\footnotesize\textsf{-2.0}} & {\footnotesize\textsf{-0.3}} & {\footnotesize\textsf{-0.3}} & {\footnotesize\textsf{-0.9}}\tabularnewline
{\footnotesize\textsf{~~Horizontal}} & {\footnotesize\textsf{897.4}} & {\footnotesize\textsf{98.0}} & {\footnotesize\textsf{28.8}} & {\footnotesize\textsf{8.5}} & {\footnotesize\textsf{61.5}} &  & {\footnotesize\textsf{5.4}} & {\footnotesize\textsf{-14.4}} & {\footnotesize\textsf{-3.5}} & {\footnotesize\textsf{-6.0}} & {\footnotesize\textsf{-7.6}}\tabularnewline
{\footnotesize\textsf{~~Angular}} & {\footnotesize\textsf{905.3}} & {\footnotesize\textsf{97.6}} & {\footnotesize\textsf{27.8}} & {\footnotesize\textsf{8.3}} & {\footnotesize\textsf{61.1}} &  & {\footnotesize\textsf{4.6}} & {\footnotesize\textsf{-11.0}} & {\footnotesize\textsf{-0.3}} & {\footnotesize\textsf{-1.1}} & {\footnotesize\textsf{-3.9}}\tabularnewline
\midrule
\multicolumn{3}{l}{{\footnotesize\textsf{\textit{Recalibration of weighted combining}}}} &  &  &  &  &  &  &  &  & \tabularnewline
{\footnotesize\textsf{~~Vertical}} & {\footnotesize\textsf{888.1}} & {\footnotesize\textsf{69.2}} & {\footnotesize\textsf{28.1}} & {\footnotesize\textsf{8.4}} & {\footnotesize\textsf{51.6}} &  & {\footnotesize\textsf{6.4}} & {\footnotesize\textsf{-0.2}} & {\footnotesize\textsf{-1.5}} & {\footnotesize\textsf{-2.4}} & {\footnotesize\textsf{-1.2}}\tabularnewline
{\footnotesize\textsf{~~Horizontal}} & {\footnotesize\textsf{851.4}} & {\footnotesize\textsf{83.8}} & {\footnotesize\textsf{29.0}} & {\footnotesize\textsf{8.6}} & {\footnotesize\textsf{56.1}} &  & {\footnotesize\textsf{10.3}} & {\footnotesize\textsf{-10.4}} & {\footnotesize\textsf{-3.9}} & {\footnotesize\textsf{-6.4}} & {\footnotesize\textsf{-6.5}}\tabularnewline
{\footnotesize\textsf{~~Angular}} & {\footnotesize\textsf{845.4}} & {\footnotesize\textsf{69.0}} & {\footnotesize\textsf{28.1}} & {\footnotesize\textsf{8.4}} & {\footnotesize\textsf{50.7}} &  & {\footnotesize\textsf{10.9}} & {\footnotesize\textsf{0.6}} & {\footnotesize\textsf{-0.9}} & {\footnotesize\textsf{-2.8}} & {\footnotesize\textsf{-0.8}}\tabularnewline
\bottomrule
\end{tabular}

{\small\vspace{-0.6cm}
}{\small\par}
\end{table}

{\scriptsize\textsf{Note: The unit of the score is deaths. Lower }}{\scriptsize\textsf{\textcolor{blue}{scores}}}{\scriptsize\textsf{
and higher skill }}{\scriptsize\textsf{\textcolor{blue}{scores}}}{\scriptsize\textsf{
are better. Skill }}{\scriptsize\textsf{\textcolor{blue}{scores use}}}{\scriptsize\textsf{
vertical averaging from Table~\ref{Table 1: MQS} as benchmark. Bold
indicates the best method in each column.}}{\scriptsize\par}

In Table~\ref{Table 3: MQS for various methods}, angular weighted
combining performs well, with the bold font indicating that it delivered
the lowest MQS scores for four of the five categories. Turning to
the other methods, we first note that the beta-transformed linear
pool performed notably well for the U.S. national series, but apart
from this, was uncompetitive with the better of the other methods.
The best trimming method was exterior trimming with vertical averaging.
In terms of the skill score averaged over all 52 series, this method
was better than any other method, slightly outperforming angular weighted
combining. Recalibration of the weighted combining methods did not
lead to improvement in the results for these methods.

It could be suggested that angular combining can be approximated by
a \textit{secondary} weighted combination of the CDF forecasts produced
by horizontal and vertical combining, with the weights related to
the angle $\theta$. When $\theta=0^{\textrm{o}}$, the weight on
the horizontal combination would equal 1, and when $\theta=90^{\textrm{o}}$,
the weight on the vertical combination would be 1. However, the question
would then be whether to perform this secondary weighted combination
horizontally or vertically. We implemented both possibilities, with
the weights estimated by minimizing the sum of in-sample MQS. The
third and fourth rows of values in Table~\ref{Table 4: MQS for combinations of combinations}
are the out-of-sample MQS results for horizontal and vertical secondary
weighted combinations of the two CDF forecasts resulting from horizontal
and vertical averaging of the individual CDF forecasts. The better
of these two methods is not quite as accurate as angular averaging,
which is included at the top of the table, and the poorer of the two
methods is notably worse than angular averaging. In this comparison,
angular averaging has the appeal of accuracy, and avoids the need
to select between horizontal and vertical secondary weighted combining.
Similar comments can be made when comparing the angular weighted combining
results in the second row of the table, with the results in the bottom
two rows, which correspond to secondary horizontal and vertical weighted
combining of the two CDF forecasts produced by horizontal and vertical
performance-weighted combining of the individual CDF forecasts.

\begin{table}[H]
{\footnotesize\caption{\linespread{0.50}\selectfont{}\label{Table 4: MQS for combinations of combinations}For
the Covid dataset, MQS for the out-of-sample period for combinations
of combinations, as well as angular combining.}
\vspace{0.1cm}
}\linespread{0.50}\selectfont{}%
\begin{tabular}{l>{\centering}p{0.8cm}>{\centering}p{0.7cm}>{\centering}p{0.9cm}>{\centering}p{0.6cm}>{\centering}p{0.3cm}c>{\centering}p{0.6cm}>{\centering}p{0.6cm}>{\centering}p{0.9cm}>{\centering}p{0.6cm}>{\centering}p{0.4cm}}
\toprule 
 & \multicolumn{5}{c}{{\footnotesize\textsf{MQS}}} & ~~~~~ & \multicolumn{5}{c}{{\footnotesize\textsf{MQS skill score (\%)}}}\tabularnewline
 & {\footnotesize\textsf{U.S.}} & {\footnotesize\textsf{High}} & {\footnotesize\textsf{Medium}} & {\footnotesize\textsf{Low}} & {\footnotesize\textsf{All}} &  & {\footnotesize\textsf{U.S.}} & {\footnotesize\textsf{High}} & {\footnotesize\textsf{Medium}} & {\footnotesize\textsf{Low}} & {\footnotesize\textsf{All}}\tabularnewline
\midrule
\multicolumn{12}{l}{{\footnotesize\textsf{\textit{Angular combining (from Table \ref{Table 1: MQS})}}}}\tabularnewline
{\footnotesize\textsf{~~Angular averaging}} & {\footnotesize\textsf{903.3}} & {\footnotesize\textsf{68.5}} & {\footnotesize\textsf{27.5}} & {\footnotesize\textsf{8.2}} & {\footnotesize\textsf{51.5}} &  & {\footnotesize\textsf{4.8}} & {\footnotesize\textsf{1.0}} & {\footnotesize\textsf{0.9}} & {\footnotesize\textsf{0.1}} & {\footnotesize\textsf{0.8}}\tabularnewline
{\footnotesize\textsf{~~Angular weighted combining}} & {\footnotesize\textsf{\textbf{839.1}}} & {\footnotesize\textsf{\textbf{66.6}}} & {\footnotesize\textsf{\textbf{27.2}}} & {\footnotesize\textsf{\textbf{8.1}}} & {\footnotesize\textsf{\textbf{49.4}}} &  & {\footnotesize\textsf{\textbf{11.6}}} & {\footnotesize\textsf{\textbf{4.0}}} & {\footnotesize\textsf{\textbf{2.4}}} & {\footnotesize\textsf{\textbf{1.1}}} & {\footnotesize\textsf{\textbf{2.7}}}\tabularnewline
\midrule
{\footnotesize\textsf{\textit{Combining horizontal \& vertical averages}}} &  &  &  &  &  &  &  &  &  &  & \tabularnewline
{\footnotesize\textsf{~~Vertical secondary weighted combining}} & {\footnotesize\textsf{914.0}} & {\footnotesize\textsf{76.3}} & {\footnotesize\textsf{28.1}} & {\footnotesize\textsf{8.2}} & {\footnotesize\textsf{54.4}} &  & {\footnotesize\textsf{3.7}} & {\footnotesize\textsf{-6.3}} & {\footnotesize\textsf{-1.4}} & {\footnotesize\textsf{-0.3}} & {\footnotesize\textsf{-2.5}}\tabularnewline
{\footnotesize\textsf{~~Horizontal secondary weighted combining}} & {\footnotesize\textsf{901.9}} & {\footnotesize\textsf{69.9}} & {\footnotesize\textsf{27.7}} & {\footnotesize\textsf{8.2}} & {\footnotesize\textsf{51.9}} &  & {\footnotesize\textsf{5.0}} & {\footnotesize\textsf{-1.3}} & {\footnotesize\textsf{0.3}} & {\footnotesize\textsf{-0.4}} & {\footnotesize\textsf{-0.4}}\tabularnewline
\midrule 
\multicolumn{12}{l}{{\footnotesize\textsf{\textit{Combining horizontal \& vertical weighted
combinations}}}}\tabularnewline
{\footnotesize\textsf{~~Vertical secondary weighted combining}} & {\footnotesize\textsf{846.9}} & {\footnotesize\textsf{73.6}} & {\footnotesize\textsf{27.4}} & {\footnotesize\textsf{\textbf{8.1}}} & {\footnotesize\textsf{52.0}} &  & {\footnotesize\textsf{10.8}} & {\footnotesize\textsf{-2.4}} & {\footnotesize\textsf{1.8}} & {\footnotesize\textsf{0.4}} & {\footnotesize\textsf{0.2}}\tabularnewline
{\footnotesize\textsf{~~Horizontal secondary weighted combining}} & {\footnotesize\textsf{839.8}} & {\footnotesize\textsf{67.3}} & {\footnotesize\textsf{27.4}} & {\footnotesize\textsf{\textbf{8.1}}} & {\footnotesize\textsf{49.7}} &  & {\footnotesize\textsf{11.5}} & {\footnotesize\textsf{2.8}} & {\footnotesize\textsf{1.9}} & {\footnotesize\textsf{0.3}} & {\footnotesize\textsf{1.9}}\tabularnewline
\bottomrule
\end{tabular}

{\small\vspace{-0.6cm}
}{\small\par}
\end{table}

{\scriptsize\textsf{Note: The unit of the score is deaths. Lower }}{\scriptsize\textsf{\textcolor{blue}{scores}}}{\scriptsize\textsf{
and higher skill }}{\scriptsize\textsf{\textcolor{blue}{scores}}}{\scriptsize\textsf{
are better. Skill }}{\scriptsize\textsf{\textcolor{blue}{scores use}}}{\scriptsize\textsf{
vertical averaging from Table~\ref{Table 1: MQS} as benchmark. Bold
indicates the best method in each column. }}{\scriptsize\par}

\section{Additional Empirical Studies\label{Section 6}}

\textcolor{blue}{We now broaden our empirical analysis by considering
data from surveys of professional economic forecasters in Section~\ref{Section 6.1},
and electricity price data in Section~\ref{Section 6.2}. The survey
data is similar to the Covid dataset in the sense that the probabilistic
forecasts are provided by a sizable number of experts using a wide
variety of methods. By contrast, in our study of electricity prices,
we combine a relatively small set of probabilistic forecasts, produced
using several time series models. A further notable aspect of our
analysis of electricity prices is that it involves a much longer dataset
than our other two studies, enabling more accurate estimation of combining
method parameters, and a clearer ranking of methods, supported by
statistical testing to compare forecast accuracy. }

\subsection{Surveys of Professional Forecasters\label{Section 6.1}}

\citet{lichtendahl2013better} compare horizontal and vertical averaging
using probabilistic forecasts from the Survey of Professional Forecasters
(SPF) of the Federal Reserve Bank of Philadelphia. Weighted combining
is not considered for this data because there is not a sufficient
record of historical accuracy for each forecaster. Each quarter, participants
submit probabilistic forecasts of annual U.S. GDP growth and inflation.
We follow \citet{lichtendahl2013better} in considering the forecasts
made in each quarter for growth and inflation in \textcolor{blue}{the
same calendar} year. We \textcolor{blue}{use} growth forecasts from
the first quarter of 1982 to the fourth quarter of 2023, amounting
to 168 quarters, and inflation forecasts from the fourth quarter of
1968 to the fourth quarter of 2023, which amounted to 221 quarters.
For growth, the number of forecasters varied from 7 to 52, with a
median of 32, and for inflation, it varied from 7 to 114, with a median
of 34. 

The U.S. SPF forecasts are also used by \citet{Jose2014} to evaluate
their aggregation methods. In addition, they consider forecasts from
the SPF of the European Central Bank (ECB). We do the same, and consider
forecasts made in each quarter for GDP growth and inflation in \textcolor{blue}{the
same calendar} year, and also for unemployment, defined as the average
of the monthly unemployment for \textcolor{blue}{the calendar} year.
Forecasts for the three variables were available from the first quarter
of 1999 to the fourth quarter of 2023, which constituted 100 quarters.
For growth, the number of forecasters had a median of 51, and varied
between 40 and 65. It was similar for inflation and unemployment.

The probabilistic forecasts from both surveys are submitted in the
form of a probability for each of a set of bins specified in the survey.
To obtain a continuous CDF, like \citet{lichtendahl2013better} and
\citet{Jose2014}, we fitted piecewise-linear CDFs to the bin probabilities.
As the number of periods in our SPF datasets were of a similar order
of magnitude to our Covid dataset, for consistency, we used \textcolor{blue}{a
similar length period}\textcolor{red}{{} }for the initial in-sample
estimation of \textcolor{blue}{combining} method parameters, and then
expanded the in-sample period as we stepped forward. \textcolor{blue}{With
the focus being to forecast the economic variables for each calendar
year, combining method parameters can only be re-estimated at the
end of each year, as new observations only become available at that
time. Allowing for this, and setting the initial estimation period
to be at least 10 periods, led us to use an initial estimation period
of 13 quarters for U.S. inflation, and 12 quarters for the other four
series.}\textcolor{red}{{} }In this way, an out-of-sample forecast was
produced from \textit{n}-\textit{\textcolor{blue}{m}} forecast origins,
where \textit{n} \textcolor{blue}{and }\textit{\textcolor{blue}{m}}\textcolor{blue}{{}
are} the \textcolor{blue}{numbers} of quarters in the dataset\textcolor{red}{{}
}\textcolor{blue}{and initial estimation period, respectively}. We
optimized parameters using the in-sample CRPS. For the actual observed
growth, inflation and unemployment, we followed \citet{Clements2018}
in using the value for year \textit{T}, released in the first quarter
of year \textit{T}+1.

\begin{figure}[H]
\begin{centering}
\caption{\linespread{0.50}\selectfont{}\label{Fig: SPFAngleTimeSeriesAvg}For
the SPF datasets and angular averaging, the optimized angle $\theta$
at each forecast origin for the five series.}
\vspace{-0.2cm}
\includegraphics[scale=0.75]{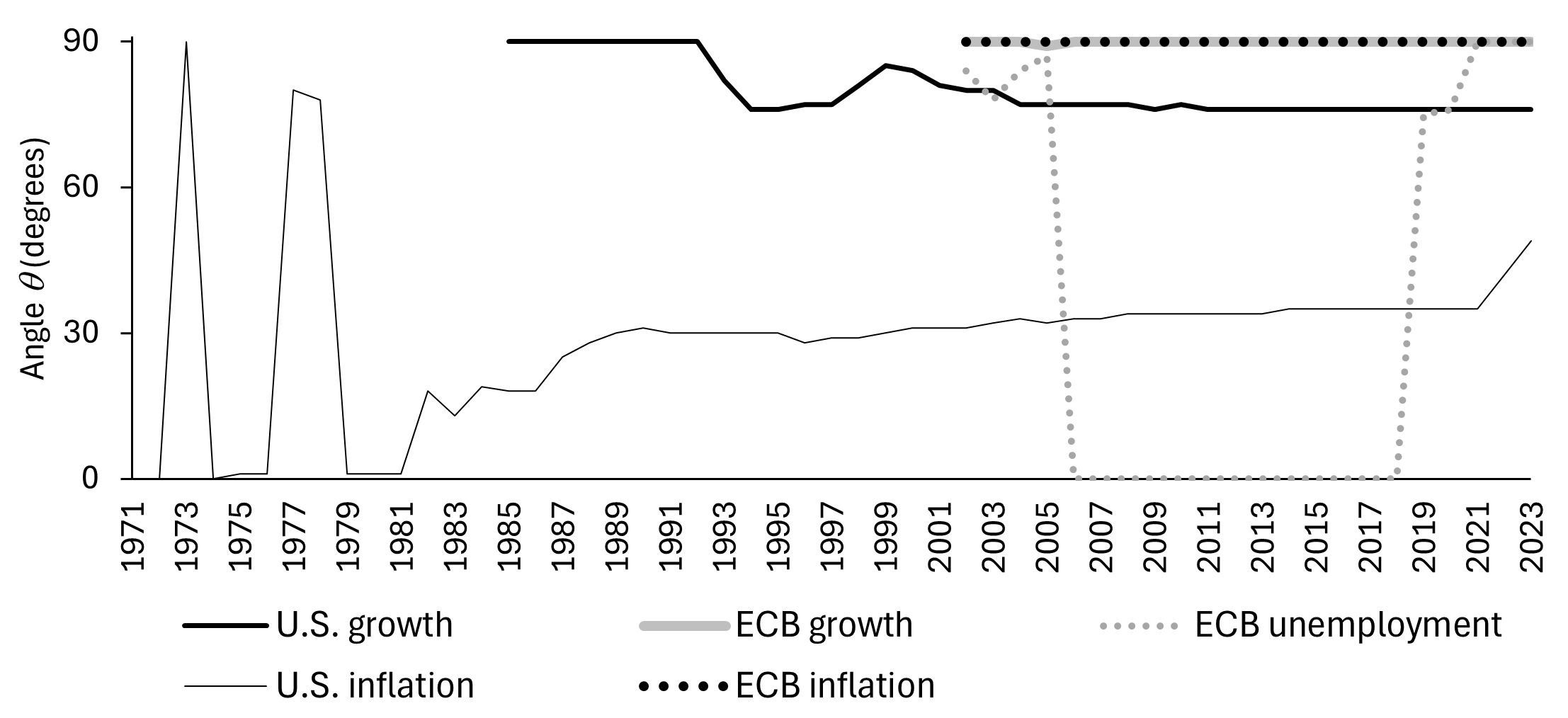}
\par\end{centering}
\centering{}\vspace{-0.7cm}
\end{figure}

Figure~\ref{Fig: SPFAngleTimeSeriesAvg} shows the angle optimized
at each forecast origin for each series. For U.S. growth, the angle
did not vary greatly, starting at $90^{\textrm{o}}$ and then remaining
at about $75^{\textrm{o}}$ for most of the forecast origins. For
U.S. inflation, the fluctuations in the angle for the early forecast
origins suggests that a larger sample size would have been better.
After the initial instability, the angle steadily rose from $0^{\textrm{o}}$
to about $50^{\textrm{o}}$. For ECB growth and inflation, the angle
was $90^{\textrm{o}}$ for almost all forecast origins, and for ECB
unemployment, the angle was $0^{\textrm{o}}$ or $90^{\textrm{o}}$
for about three-quarters of the forecast origins.

Table~\ref{Table: SPF-CRPS-Averaging} presents the out-of-sample
CRPS results for horizontal, vertical and angular averaging, along
with switching between horizontal and vertical averaging. The table
shows that angular averaging outperformed the other methods for the
two U.S. series. In view of the optimized angles in Figure~\ref{Fig: SPFAngleTimeSeriesAvg},
it is understandable that the results for angular averaging were the
same as vertical averaging for ECB growth and inflation. ECB unemployment
was the only one of the five series for which angular averaging was
outperformed. Online Appendix E.1 provides a comparison of methods
in terms of the quantile skill scores for different quantile probability
levels. These results show horizontal and angular averaging performing
particularly well in the tails of the distributions. In Online Appendix
E.2, the CRPS results for other methods show that the best results
overall were obtained using exterior trimming with angular averaging. 

\begin{table}[H]
{\footnotesize\caption{\linespread{0.50}\selectfont{}\label{Table: SPF-CRPS-Averaging}For
the SPF datasets, CRPS for the out-of-sample period for the three
forms of averaging and horizontal/vertical switching.}
\vspace{0.1cm}
}\linespread{0.50}\selectfont{}%
\begin{tabular}{lccccccccccccc}
\toprule 
 & \multicolumn{6}{c}{{\footnotesize\textsf{CRPS}}} & ~~~ & \multicolumn{6}{c}{{\footnotesize\textsf{CRPS skill score (\%)}}}\tabularnewline
 & \begin{cellvarwidth}[t]
\centering
{\footnotesize\textsf{U.S.}}{\footnotesize\par}

{\footnotesize\textsf{Gro\textsubscript{}}}
\end{cellvarwidth} & \begin{cellvarwidth}[t]
\centering
{\footnotesize\textsf{U.S.}}{\footnotesize\par}

{\footnotesize\textsf{Infl}}
\end{cellvarwidth} & \begin{cellvarwidth}[t]
\centering
{\footnotesize\textsf{ECB}}{\footnotesize\par}

{\footnotesize\textsf{Gro}}
\end{cellvarwidth} & \begin{cellvarwidth}[t]
\centering
{\footnotesize\textsf{ECB}}{\footnotesize\par}

{\footnotesize\textsf{Infl}}
\end{cellvarwidth} & \begin{cellvarwidth}[t]
\centering
{\footnotesize\textsf{ECB}}{\footnotesize\par}

{\footnotesize\textsf{Unem}}
\end{cellvarwidth} & \begin{cellvarwidth}[t]
\centering
~

{\footnotesize\textsf{All}}
\end{cellvarwidth} &  & \begin{cellvarwidth}[t]
\centering
{\footnotesize\textsf{U.S.}}{\footnotesize\par}

{\footnotesize\textsf{Gro}}
\end{cellvarwidth} & \begin{cellvarwidth}[t]
\centering
{\footnotesize\textsf{U.S.}}{\footnotesize\par}

{\footnotesize\textsf{Infl}}
\end{cellvarwidth} & \begin{cellvarwidth}[t]
\centering
{\footnotesize\textsf{ECB}}{\footnotesize\par}

{\footnotesize\textsf{Gro}}
\end{cellvarwidth} & \begin{cellvarwidth}[t]
\centering
{\footnotesize\textsf{ECB}}{\footnotesize\par}

{\footnotesize\textsf{Infl}}
\end{cellvarwidth} & \begin{cellvarwidth}[t]
\centering
{\footnotesize\textsf{ECB}}{\footnotesize\par}

{\footnotesize\textsf{Unem}}
\end{cellvarwidth} & \begin{cellvarwidth}[t]
\centering
~

{\footnotesize\textsf{All}}
\end{cellvarwidth}\tabularnewline
\midrule 
{\footnotesize\textsf{Vertical averaging}} & {\footnotesize\textsf{\textcolor{blue}{38.0}}} & {\footnotesize\textsf{\textcolor{blue}{36.4}}} & {\footnotesize\textsf{\textbf{\textcolor{blue}{80.4}}}} & {\footnotesize\textsf{\textbf{\textcolor{blue}{66.9}}}} & {\footnotesize\textsf{\textbf{\textcolor{blue}{22.6}}}} & {\footnotesize\textsf{\textcolor{blue}{48.8}}} &  & {\footnotesize\textsf{\textcolor{blue}{0.0}}} & {\footnotesize\textsf{\textcolor{blue}{0.0}}} & {\footnotesize\textsf{\textbf{\textcolor{blue}{0.0}}}} & {\footnotesize\textsf{\textbf{\textcolor{blue}{0.0}}}} & {\footnotesize\textsf{\textbf{\textcolor{blue}{0.0}}}} & {\footnotesize\textsf{\textcolor{blue}{0.0}}}\tabularnewline
{\footnotesize\textsf{Horizontal averaging}} & {\footnotesize\textsf{\textcolor{blue}{38.1}}} & {\footnotesize\textsf{\textcolor{blue}{35.9}}} & {\footnotesize\textsf{\textcolor{blue}{81.4}}} & {\footnotesize\textsf{\textcolor{blue}{67.9}}} & {\footnotesize\textsf{\textcolor{blue}{23.2}}} & {\footnotesize\textsf{\textcolor{blue}{49.3}}} &  & {\footnotesize\textsf{\textcolor{blue}{-0.2}}} & {\footnotesize\textsf{\textcolor{blue}{1.4}}} & {\footnotesize\textsf{\textcolor{blue}{-1.3}}} & {\footnotesize\textsf{\textcolor{blue}{-1.6}}} & {\footnotesize\textsf{\textcolor{blue}{-2.9}}} & {\footnotesize\textsf{\textcolor{blue}{-0.9}}}\tabularnewline
{\footnotesize\textsf{Horizontal/vertical switching}} & {\footnotesize\textsf{\textcolor{blue}{38.0}}} & {\footnotesize\textsf{\textcolor{blue}{36.1}}} & {\footnotesize\textsf{\textbf{\textcolor{blue}{80.4}}}} & {\footnotesize\textsf{\textbf{\textcolor{blue}{66.9}}}} & {\footnotesize\textsf{\textbf{\textcolor{blue}{22.6}}}} & {\footnotesize\textsf{\textcolor{blue}{48.8}}} &  & {\footnotesize\textsf{\textcolor{blue}{0.0}}} & {\footnotesize\textsf{\textcolor{blue}{0.9}}} & {\footnotesize\textsf{\textbf{\textcolor{blue}{0.0}}}} & {\footnotesize\textsf{\textbf{\textcolor{blue}{0.0}}}} & {\footnotesize\textsf{\textcolor{blue}{-0.2}}} & {\footnotesize\textsf{\textcolor{blue}{0.2}}}\tabularnewline
{\footnotesize\textsf{Angular averaging}} & {\footnotesize\textsf{\textbf{\textcolor{blue}{37.8}}}} & {\footnotesize\textsf{\textbf{\textcolor{blue}{35.8}}}} & {\footnotesize\textsf{\textbf{\textcolor{blue}{80.4}}}} & {\footnotesize\textsf{\textbf{\textcolor{blue}{66.9}}}} & {\footnotesize\textsf{\textcolor{blue}{22.8}}} & {\footnotesize\textsf{\textbf{\textcolor{blue}{48.7}}}} &  & {\footnotesize\textsf{\textbf{\textcolor{blue}{0.6}}}} & {\footnotesize\textsf{\textbf{\textcolor{blue}{1.6}}}} & {\footnotesize\textsf{\textbf{\textcolor{blue}{0.0}}}} & {\footnotesize\textsf{\textbf{\textcolor{blue}{0.0}}}} & {\footnotesize\textsf{\textcolor{blue}{-0.9}}} & {\footnotesize\textsf{\textbf{\textcolor{blue}{0.3}}}}\tabularnewline
\bottomrule
\end{tabular}

{\small\vspace{-0.6cm}
}{\small\par}
\end{table}

{\scriptsize\textsf{Note: }}{\scriptsize\textsf{\textcolor{blue}{Scores
are percentages and have been multiplied by }}}{\scriptsize\textsf{100.
Lower }}{\scriptsize\textsf{\textcolor{blue}{scores}}}{\scriptsize\textsf{
and higher skill }}{\scriptsize\textsf{\textcolor{blue}{scores}}}{\scriptsize\textsf{
are better. Skill }}{\scriptsize\textsf{\textcolor{blue}{scores use}}}{\scriptsize\textsf{
vertical averaging as benchmark. Bold indicates the best method in
each column.}}{\scriptsize\par}

\subsection{Electricity Prices\label{Section 6.2}}

\textcolor{blue}{Our third empirical study uses} hourly Nord Pool
day-ahead electricity market clearing prices. The price for each hour
is set the previous day, and this has led to models being fitted separately
to the daily time series for each hour. We focused on day-ahead forecasting,
and implemented the following methods for each of the 24 hourly time
series:

\textit{Method 1. Historical simulation} -- As a simple benchmark,
we used the piecewise-linear CDF constructed from the empirical distribution
of the price for the same hour on the previous 28 days. Optimizing
the number of previous days to use in the method led to fewer than
10, which we felt was unappealing as the basis for estimating a CDF,
so we arbitrarily chose a period of one month. 

\textit{Method 2. ARX} -- We fitted the following model considered
by \citet{Nowotarski2018},{\small\vspace{-0.5cm}
}
\begin{equation}
p_{h,t}=\phi_{h,0}+\phi_{h,1}p_{h,t-1}+\phi_{h,2}p_{h,t-2}+\phi_{h,3}p_{h,t-7}+\phi_{h,4}mp_{t-1}+d_{h,1}D_{Mon,t}+d_{h,2}D_{Sat,t}+d_{h,3}D_{Sun,t}+\varepsilon_{h,t}\label{eq:ARX}
\end{equation}
where $p_{h,t}$ is the log price for hour $h$ on day $t$; $mp_{t}$
is the minimum value of the log price on day $t$; $D_{Mon,t}$, $D_{Sat,t}$
and $D_{Sun,t}$ are binary variables indicating whether day $t$
is a Monday, Saturday or Sunday, respectively; the $\phi_{h,i}$ and
$d_{h,i}$ are parameters; and the $\varepsilon_{h,t}$ is an error
term. We estimated the model using least squares. A CDF forecast was
then constructed with the model's point forecast as the mean, and
distribution around the mean obtained by applying the historical simulation
approach to the last 28 residuals from the model, with the aim of
capturing time-varying volatility. 

\textit{Method 3. AR1-GARCH} -- This method has an AR(1) model for
the mean of the log price, and the asymmetric GJR-GARCH(1,1) model
for the variance. We estimated the model using maximum likelihood
based on the skewed t distribution described by \citet[Chapter 6, Section 7]{christoffersen2011elements}.
To construct each CDF forecast, we used the forecast of the mean and
variance with distributional shape specified as the piecewise-linear
CDF constructed from the empirical distribution of all standardized
residuals.

\textit{Method 4. ARX-GARCH} -- This is identical to the AR1-GARCH
method, except that the mean was modeled using (\ref{eq:ARX}). This
model was fitted by \citet{Taylor2021} to the same electricity price
data.

Our dataset consisted of the 6$\times$365 days starting on January
1 2013. Online Appendix F.1 presents the time series plot of the price
for one hour of the day. \textcolor{blue}{As the time series were
much longer than those in our other two empirical studies, we used
a rolling window for estimation, which is more common than an expanding
window when a reasonably long time series is available. An advantage
of a rolling window is that it enables the straightforward application
of Diebold-Mariano tests to compare forecast accuracy (\citealt{Giacomini2006}).}
For each hour \textcolor{blue}{of our electricity price time series},
we used a rolling window of 365 days for repeated re-estimation of
the parameters in Methods 2 to 4. For all four methods, we produced
out-of-sample day-ahead forecasts for the final 5$\times$365 days.
Using this period, we repeatedly re-estimated combining method parameters
by minimizing the CRPS based on a rolling window of 365 days. This
resulted in out-of-sample forecasts for the combining methods for
the final 4$\times$365 days. Given the availability of individual
model forecasts for the sizable in-sample period, we adapted the approach
to computing combining weights that we used in our Covid study, so
that the weight on each method was inversely proportional to the method's
CRPS raised to the power of a tuning parameter (see, for example,
\citealt{Stock1999}). This parameter was estimated for each weighted
combining method at each forecast origin.\textcolor{blue}{{} }

As the features of Methods 2 and 3 are captured by Method 4, it seemed
likely this model would dominate weighted combinations. Therefore,
we considered combinations involving three different selections of
the four methods: Methods 1 and 2, Methods 1 to 3, and Methods 1 to
4.

\begin{figure}[H]
\begin{centering}
\caption{\linespread{0.50}\selectfont{}\label{Fig: ElecAngleHistogramAvgOf3}For
the electricity price dataset and angular averaging of Methods 1 to
3, histogram of the angle $\theta$ optimized at each forecast origin
for all 24 hourly series.}
\vspace{-0.2cm}
\includegraphics[scale=0.8]{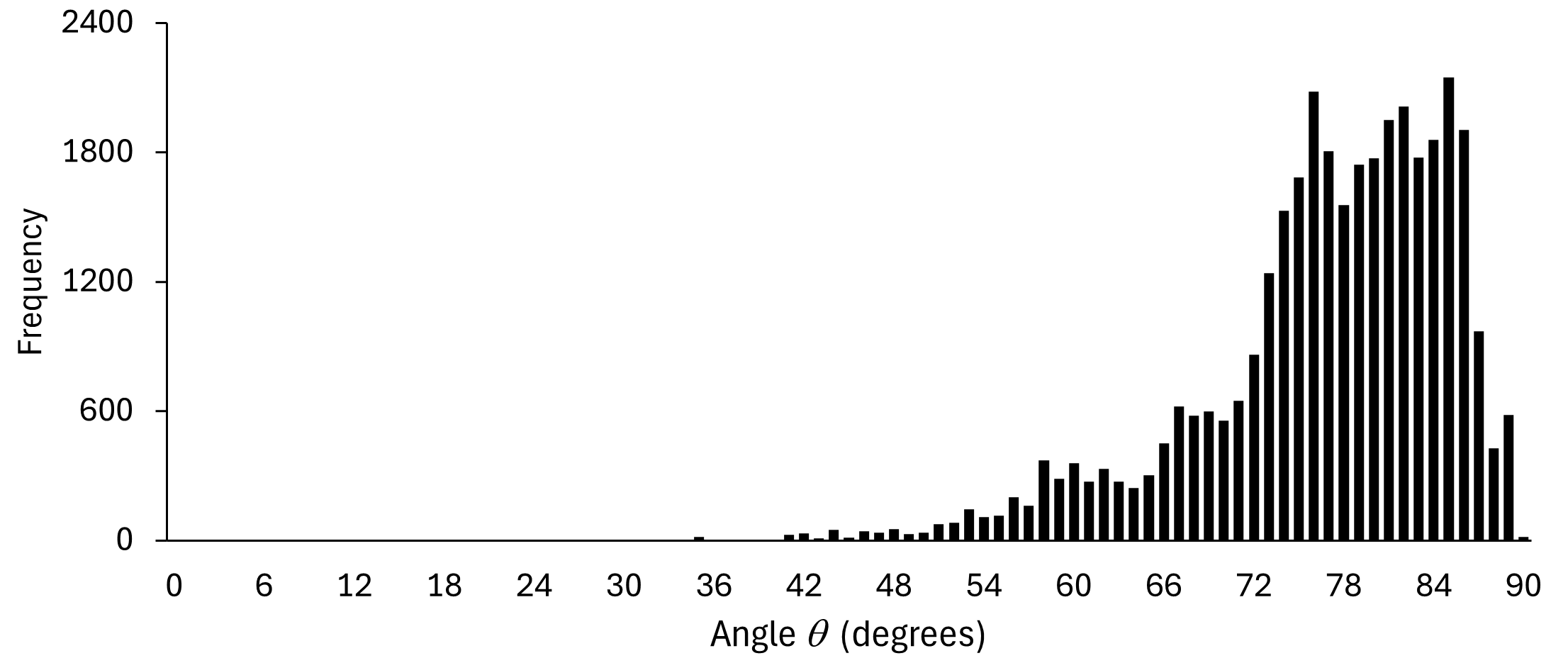}
\par\end{centering}
\centering{}\vspace{-0.8cm}
\end{figure}

\begin{figure}[H]
\begin{centering}
\caption{\linespread{0.50}\selectfont{}\label{Fig: ElecAngleTimeSeriesAvgOf3}For
the electricity price dataset and angular averaging of Methods 1 to
3, the optimized angle $\theta$ at each forecast origin for four
of the hourly series.}
\vspace{-0.2cm}
\includegraphics[scale=0.75]{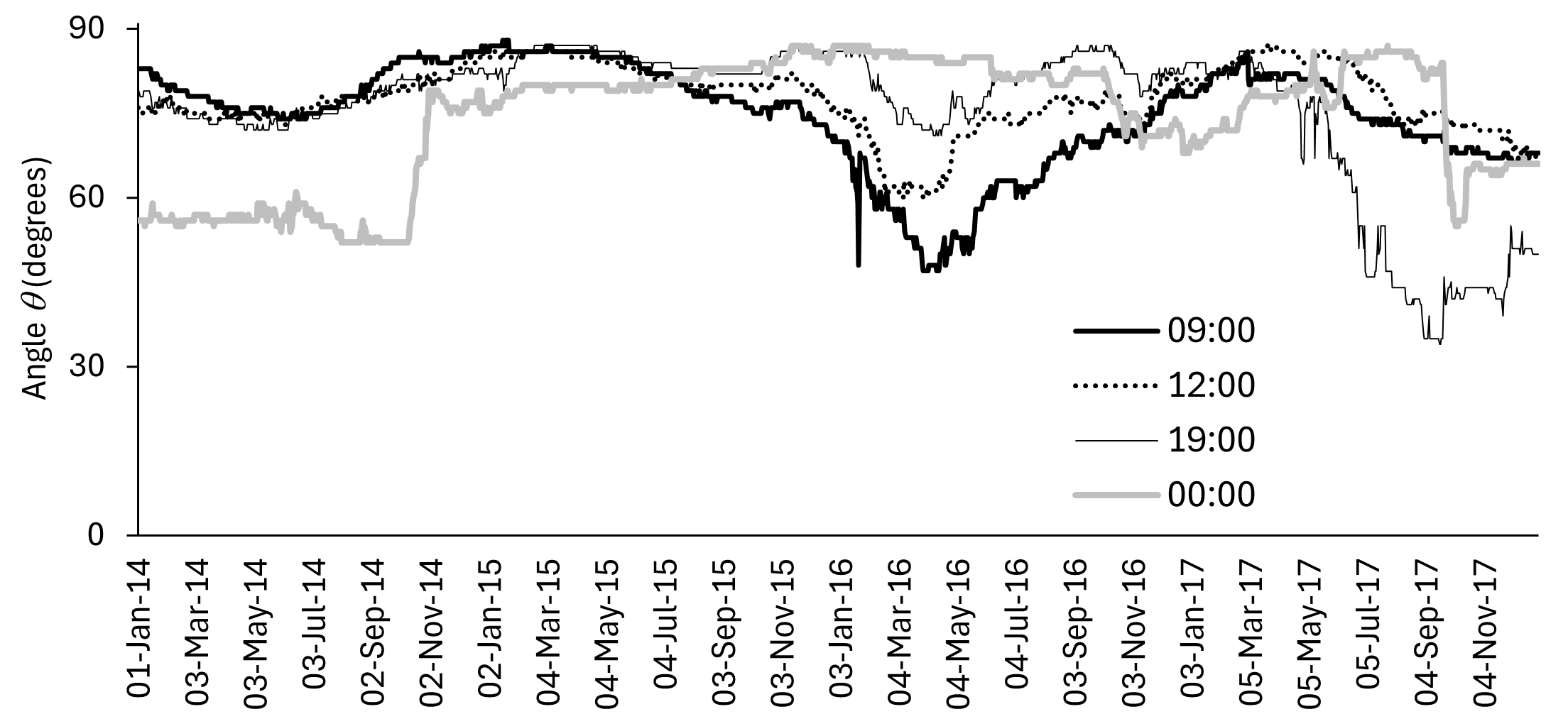}
\par\end{centering}
\centering{}\vspace{-0.7cm}
\end{figure}

For angular averaging of Methods 1 to 3, Figure~\ref{Fig: ElecAngleHistogramAvgOf3}
presents the histogram of the angles optimized at all 4$\times$365
forecast origins for all 24 hourly series. Figure~\ref{Fig: ElecAngleTimeSeriesAvgOf3}
shows the angle optimized at each forecast origin for the following
four hourly series: midday, midnight\textcolor{blue}{,} and the two
hours in the day corresponding to the highest values of the median
of the price for the entire sample of 6$\times$365 days. Each series
shows the angle steadily varying over time, fluctuating by about $30^{\textrm{o}}$
between its minimum and maximum.

Table~\ref{Table: ELEC-CRPS-AvgWtdCombSmry} shows the CRPS averaged
across the out-of-sample period for all 24 hourly series for combining
applied to the three selections of Methods 1 to 4. Comparing the methods
involving averaging, horizontal averaging was the least accurate,
the switching method performed similarly to vertical averaging, and
angular averaging was the most accurate. We note that the angular
average of Methods 1 to 3, and of Methods 1 to 4, was more accurate
than the best of the four individual methods. The table shows weighted
combining outperforming averaging. Of the weighted combining methods,
angular combining performed the best. However, the difference between
the best and worst weighted combining results was notably less than
for the averaging methods.

\begin{table}[H]
{\footnotesize\caption{\linespread{0.50}\selectfont{}\label{Table: ELEC-CRPS-AvgWtdCombSmry}For
the electricity price dataset, CRPS and CRPS skill score for the out-of-sample
period of 4$\times$365 days, averaged across all 24 hourly series
for the four individual methods, combining methods, and horizontal/vertical
switching applied to Methods 1 and 2, Methods 1 to 3, and Methods
1 to 4.}
\vspace{0.1cm}
}\linespread{0.50}\selectfont{}%
\begin{tabular}{lcccccccc}
\toprule 
 & \multicolumn{2}{c}{{\footnotesize\textsf{Methods 1 \& 2}}} & ~~ & \multicolumn{2}{c}{{\footnotesize\textsf{Methods 1 to 3}}} & ~~ & \multicolumn{2}{c}{{\footnotesize\textsf{Methods 1 to 4}}}\tabularnewline
 & {\footnotesize\textsf{CRPS}} & {\footnotesize\textsf{Skill score (\%)}} &  & {\footnotesize\textsf{CRPS}} & {\footnotesize\textsf{Skill score (\%)}} &  & {\footnotesize\textsf{CRPS}} & {\footnotesize\textsf{Skill score (\%)}}\tabularnewline
\midrule
{\footnotesize\textsf{\textit{Averaging}}} &  &  &  &  &  &  &  & \tabularnewline
{\footnotesize\textsf{~~Vertical}} & {\footnotesize\textsf{187.1}} & {\footnotesize\textsf{0.0}} &  & {\footnotesize\textsf{178.9}} & {\footnotesize\textsf{0.0}} &  & {\footnotesize\textsf{173.1}} & {\footnotesize\textsf{0.0}}\tabularnewline
{\footnotesize\textsf{~~Horizontal}} & {\footnotesize\textsf{191.6}} & {\footnotesize\textsf{-2.6}} &  & {\footnotesize\textsf{181.1}} & {\footnotesize\textsf{-1.4}} &  & {\footnotesize\textsf{174.9}} & {\footnotesize\textsf{-1.2}}\tabularnewline
{\footnotesize\textsf{~~Horizontal/vertical switching}} & {\footnotesize\textsf{187.3}} & {\footnotesize\textsf{-0.1}} &  & {\footnotesize\textsf{178.9}} & {\footnotesize\textsf{0.0}} &  & {\footnotesize\textsf{173.1}} & {\footnotesize\textsf{0.0}}\tabularnewline
{\footnotesize\textsf{~~Angular}} & {\footnotesize\textsf{186.0}} & {\footnotesize\textsf{0.6}} &  & {\footnotesize\textsf{177.7}} & {\footnotesize\textsf{0.6}} &  & {\footnotesize\textsf{172.2}} & {\footnotesize\textsf{0.4}}\tabularnewline
\midrule 
{\footnotesize\textsf{\textit{Weighted combining}}} &  &  &  &  &  &  &  & \tabularnewline
{\footnotesize\textsf{~~Vertical}} & {\footnotesize\textsf{176.4}} & {\footnotesize\textsf{6.1}} &  & {\footnotesize\textsf{171.9}} & {\footnotesize\textsf{3.9}} &  & {\footnotesize\textsf{169.8}} & {\footnotesize\textsf{1.9}}\tabularnewline
{\footnotesize\textsf{~~Horizontal}} & {\footnotesize\textsf{177.9}} & {\footnotesize\textsf{5.3}} &  & {\footnotesize\textsf{171.4}} & {\footnotesize\textsf{4.1}} &  & {\footnotesize\textsf{170.0}} & {\footnotesize\textsf{1.8}}\tabularnewline
{\footnotesize\textsf{~~Horizontal/vertical switching}} & {\footnotesize\textsf{176.5}} & {\footnotesize\textsf{6.0}} &  & {\footnotesize\textsf{171.5}} & {\footnotesize\textsf{4.1}} &  & {\footnotesize\textsf{169.8}} & {\footnotesize\textsf{1.9}}\tabularnewline
{\footnotesize\textsf{~~Angular}} & {\footnotesize\textsf{\textbf{175.6}}} & {\footnotesize\textsf{\textbf{6.5}}} &  & {\footnotesize\textsf{\textbf{170.7}}} & {\footnotesize\textsf{\textbf{4.5}}} &  & {\footnotesize\textsf{\textbf{169.4}}} & {\footnotesize\textsf{\textbf{2.1}}}\tabularnewline
\midrule
{\footnotesize\textsf{\textit{Individual methods}}} &  &  &  &  &  &  &  & \tabularnewline
{\footnotesize\textsf{~~1. Historical simulation}} & {\footnotesize\textsf{267.1}} & {\footnotesize\textsf{-43.6}} &  & {\footnotesize\textsf{267.1}} & {\footnotesize\textsf{-51.5}} &  & {\footnotesize\textsf{267.1}} & {\footnotesize\textsf{-56.7}}\tabularnewline
{\footnotesize\textsf{~~2. ARX}} & {\footnotesize\textsf{182.4}} & {\footnotesize\textsf{3.1}} &  & {\footnotesize\textsf{182.4}} & {\footnotesize\textsf{-2.3}} &  & {\footnotesize\textsf{182.4}} & {\footnotesize\textsf{-5.7}}\tabularnewline
{\footnotesize\textsf{~~3. AR1-GARCH}} & {\footnotesize\textsf{NA}} & {\footnotesize\textsf{NA}} &  & {\footnotesize\textsf{198.6}} & {\footnotesize\textsf{-8.9}} &  & {\footnotesize\textsf{198.6}} & {\footnotesize\textsf{-12.6}}\tabularnewline
{\footnotesize\textsf{~~4. ARX-GARCH}} & {\footnotesize\textsf{NA}} & {\footnotesize\textsf{NA}} &  & {\footnotesize\textsf{NA}} & {\footnotesize\textsf{NA}} &  & {\footnotesize\textsf{172.8}} & {\footnotesize\textsf{0.4}}\tabularnewline
\bottomrule
\end{tabular}

{\small\vspace{-0.6cm}
}{\small\par}
\end{table}

{\scriptsize\textsf{Note: }}{\scriptsize\textsf{\textcolor{blue}{Scores
have}}}{\scriptsize\textsf{ unit EUR/MWh }}{\scriptsize\textsf{\textcolor{blue}{and
have been multiplied by }}}{\scriptsize\textsf{100. Lower}}{\scriptsize\textsf{\textcolor{blue}{{}
scores}}}{\scriptsize\textsf{ and higher skill }}{\scriptsize\textsf{\textcolor{blue}{scores}}}{\scriptsize\textsf{
are better. Skill }}{\scriptsize\textsf{\textcolor{blue}{scores use}}}{\scriptsize\textsf{
vertical averaging as benchmark. NA abbreviates ``not applicable''.
Bold indicates the best method in each column.}}{\scriptsize\par}

As this study involved a lengthy out-of-sample period, we performed
statistical testing to compare the CRPS results. For combinations
of Methods 1 to 3, Table~\ref{Table: ELEC-DieboldMariano} shows
the number of hourly series, out of the 24, for which the Diebold-Mariano
test indicated the CRPS for the method listed in the row was significantly
lower than for the method listed in the column (using a 5\% significance
level). The row for angular averaging shows that it was significantly
more accurate than the other averaging methods for more than half
the 24 series. The column for the angular average shows that it was
not significantly less accurate than the other averaging methods for
any series. The row for angular weighted combining shows that it was
significantly more accurate than the other weighted combining methods
for varying numbers of the 24 series, and significantly less accurate
than none of them. Online Appendix F.2 shows similar results for combinations
of Methods 1 and 2, and of Methods 1 to 4.

\begin{table}[H]
{\footnotesize\caption{\linespread{0.50}\selectfont{}\label{Table: ELEC-DieboldMariano}For
the electricity price dataset, Diebold-Mariano test results for combinations
of Methods 1 to 3. Values are the number of hourly series, out of
the 24, for which the CRPS for the method listed in the row was significantly
lower than for the method listed in the column (using a 5\% significance
level).}
\vspace{0.1cm}
}\linespread{0.50}\selectfont{}%
\begin{tabular}{lcccccccccccccccc}
\toprule 
 & {\footnotesize\textsf{\rotatebox{90}{\textit{Averaging}}}} & {\footnotesize\textsf{\rotatebox{90}{\hspace{0.2cm}Vertical}}} & {\footnotesize\textsf{\rotatebox{90}{\hspace{0.2cm}Horizontal}}} & {\footnotesize\textsf{\rotatebox{90}{\hspace{0.2cm}Hor/vert switch}}} & {\footnotesize\textsf{\rotatebox{90}{\hspace{0.2cm}Angular}}} & ~~~ & {\footnotesize\textsf{\rotatebox{90}{\textit{Wtd combining}}}} & {\footnotesize\textsf{\rotatebox{90}{\hspace{0.2cm}Vertical}}} & {\footnotesize\textsf{\rotatebox{90}{\hspace{0.2cm}Horizontal}}} & {\footnotesize\textsf{\rotatebox{90}{\hspace{0.2cm}Hor/vert switch}}} & {\footnotesize\textsf{\rotatebox{90}{\hspace{0.2cm}Angular}}} & ~~~ & {\footnotesize\textsf{\rotatebox{90}{\textit{Indiv methods}}}} & {\footnotesize\textsf{\rotatebox{90}{\hspace{0.2cm}1. Hist sim}}} & {\footnotesize\textsf{\rotatebox{90}{\hspace{0.2cm}2. ARX}}} & {\footnotesize\textsf{\rotatebox{90}{\hspace{0.2cm}3. AR1-GARCH}}}\tabularnewline
\midrule
{\footnotesize\textsf{\textit{Averaging}}} &  &  &  &  &  &  &  &  &  &  &  &  &  &  &  & \tabularnewline
{\footnotesize\textsf{~~Vertical}} &  & {\footnotesize\textsf{NA}} & {\footnotesize\textsf{7}} & {\footnotesize\textsf{0}} & {\footnotesize\textsf{0}} &  &  & {\footnotesize\textsf{0}} & {\footnotesize\textsf{0}} & {\footnotesize\textsf{0}} & {\footnotesize\textsf{0}} &  &  & {\footnotesize\textsf{\textbf{24}}} & {\footnotesize\textsf{5}} & {\footnotesize\textsf{16}}\tabularnewline
{\footnotesize\textsf{~~Horizontal}} &  & {\footnotesize\textsf{0}} & {\footnotesize\textsf{NA}} & {\footnotesize\textsf{0}} & {\footnotesize\textsf{0}} &  &  & {\footnotesize\textsf{0}} & {\footnotesize\textsf{0}} & {\footnotesize\textsf{0}} & {\footnotesize\textsf{0}} &  &  & {\footnotesize\textsf{\textbf{24}}} & {\footnotesize\textsf{2}} & {\footnotesize\textsf{14}}\tabularnewline
{\footnotesize\textsf{~~Horizontal/vertical switching}} &  & {\footnotesize\textsf{0}} & {\footnotesize\textsf{7}} & {\footnotesize\textsf{NA}} & {\footnotesize\textsf{0}} &  &  & {\footnotesize\textsf{0}} & {\footnotesize\textsf{0}} & {\footnotesize\textsf{0}} & {\footnotesize\textsf{0}} &  &  & {\footnotesize\textsf{\textbf{24}}} & {\footnotesize\textsf{5}} & {\footnotesize\textsf{16}}\tabularnewline
{\footnotesize\textsf{~~Angular}} &  & {\footnotesize\textsf{16}} & {\footnotesize\textsf{22}} & {\footnotesize\textsf{15}} & {\footnotesize\textsf{NA}} &  &  & {\footnotesize\textsf{0}} & {\footnotesize\textsf{0}} & {\footnotesize\textsf{0}} & {\footnotesize\textsf{0}} &  &  & {\footnotesize\textsf{\textbf{24}}} & {\footnotesize\textsf{6}} & {\footnotesize\textsf{16}}\tabularnewline
\midrule 
{\footnotesize\textsf{\textit{Weighted combining}}} &  &  &  &  &  &  &  &  &  &  &  &  &  &  &  & \tabularnewline
{\footnotesize\textsf{~~Vertical}} &  & {\footnotesize\textsf{18}} & {\footnotesize\textsf{21}} & {\footnotesize\textsf{18}} & {\footnotesize\textsf{13}} &  &  & {\footnotesize\textsf{NA}} & {\footnotesize\textsf{0}} & {\footnotesize\textsf{1}} & {\footnotesize\textsf{0}} &  &  & {\footnotesize\textsf{\textbf{24}}} & {\footnotesize\textsf{\textbf{24}}} & {\footnotesize\textsf{\textbf{23}}}\tabularnewline
{\footnotesize\textsf{~~Horizontal}} &  & {\footnotesize\textsf{16}} & {\footnotesize\textsf{21}} & {\footnotesize\textsf{16}} & {\footnotesize\textsf{14}} &  &  & {\footnotesize\textsf{2}} & {\footnotesize\textsf{NA}} & {\footnotesize\textsf{0}} & {\footnotesize\textsf{0}} &  &  & {\footnotesize\textsf{\textbf{24}}} & {\footnotesize\textsf{\textbf{24}}} & {\footnotesize\textsf{\textbf{23}}}\tabularnewline
{\footnotesize\textsf{~~Horizontal/vertical switching}} &  & {\footnotesize\textsf{19}} & {\footnotesize\textsf{23}} & {\footnotesize\textsf{20}} & {\footnotesize\textsf{14}} &  &  & {\footnotesize\textsf{2}} & {\footnotesize\textsf{0}} & {\footnotesize\textsf{NA}} & {\footnotesize\textsf{0}} &  &  & {\footnotesize\textsf{\textbf{24}}} & {\footnotesize\textsf{\textbf{24}}} & {\footnotesize\textsf{\textbf{23}}}\tabularnewline
{\footnotesize\textsf{~~Angular}} &  & {\footnotesize\textsf{\textbf{21}}} & {\footnotesize\textsf{\textbf{24}}} & {\footnotesize\textsf{\textbf{21}}} & {\footnotesize\textsf{\textbf{17}}} &  &  & {\footnotesize\textsf{\textbf{15}}} & {\footnotesize\textsf{\textbf{4}}} & {\footnotesize\textsf{\textbf{12}}} & {\footnotesize\textsf{NA}} &  &  & {\footnotesize\textsf{\textbf{24}}} & {\footnotesize\textsf{\textbf{24}}} & {\footnotesize\textsf{\textbf{23}}}\tabularnewline
\midrule
{\footnotesize\textsf{\textit{Individual methods}}} &  &  &  &  &  &  &  &  &  &  &  &  &  &  &  & \tabularnewline
{\footnotesize\textsf{~~1. Historical simulation}} &  & {\footnotesize\textsf{0}} & {\footnotesize\textsf{0}} & {\footnotesize\textsf{0}} & {\footnotesize\textsf{0}} &  &  & {\footnotesize\textsf{0}} & {\footnotesize\textsf{0}} & {\footnotesize\textsf{0}} & {\footnotesize\textsf{0}} &  &  & {\footnotesize\textsf{NA}} & {\footnotesize\textsf{0}} & {\footnotesize\textsf{0}}\tabularnewline
{\footnotesize\textsf{~~2. ARX}} &  & {\footnotesize\textsf{0}} & {\footnotesize\textsf{2}} & {\footnotesize\textsf{0}} & {\footnotesize\textsf{0}} &  &  & {\footnotesize\textsf{0}} & {\footnotesize\textsf{0}} & {\footnotesize\textsf{0}} & {\footnotesize\textsf{0}} &  &  & {\footnotesize\textsf{\textbf{24}}} & {\footnotesize\textsf{NA}} & {\footnotesize\textsf{11}}\tabularnewline
{\footnotesize\textsf{~~3. AR1-GARCH}} &  & {\footnotesize\textsf{0}} & {\footnotesize\textsf{1}} & {\footnotesize\textsf{0}} & {\footnotesize\textsf{0}} &  &  & {\footnotesize\textsf{0}} & {\footnotesize\textsf{0}} & {\footnotesize\textsf{0}} & {\footnotesize\textsf{0}} &  &  & {\footnotesize\textsf{\textbf{24}}} & {\footnotesize\textsf{4}} & {\footnotesize\textsf{NA}}\tabularnewline
\bottomrule
\end{tabular}

{\small\vspace{-0.6cm}
}{\small\par}
\end{table}

{\scriptsize\textsf{Note: NA abbreviates ``not applicable''. Bold
indicates the best method in each column.}}{\scriptsize\par}

\smallskip{}

In Online Appendix F.3, we present the CRPS for each hour of the day.
Online Appendix F.4 provides the quantile skill scores for different
quantile probability levels, and, as in the Covid and SPF studies,
angular combining can be seen to perform particularly well in the
tails of the distributions. Online Appendix F.5 evaluates other combining
methods, and shows that the results for the beta-transformed linear
pool matched those for angular weighted combining. 

\section{Summary and Concluding Remarks\label{Section 7}}

In the literature on distributional forecasting, the question has
been posed as to whether it is better to average horizontally or vertically.
We have extended the debate by proposing combining at an angle, leading
to a method that lies between the extremes of horizontal and vertical
combining. Although our proposal is motivated by empirical pragmatism,
we also feel it is interesting conceptually. We extended the theoretical
results of \citet{lichtendahl2013better} to obtain results for angular
combining. Like vertical and horizontal combining, angular combining
produces CDF forecasts for which the mean is the combination of the
means of the individual CDF forecasts. Angular combining produces
a CDF with lower variance than vertical combining, and, under certain
assumptions, greater variance than horizontal combining. As with vertical
and horizontal averaging, the CRPS for angular averaging is no worse
than the average CRPS of the individual CDF forecasts. \textcolor{blue}{In
Online Appendix G, we present} a new modeling framework that reveals
how horizontal, vertical and angular combining structure the available
information differently\textcolor{blue}{, providing insight for when
it is most natural to use each combining method, and showing potential
for new approaches to combining}.

In our empirical studies, angular combining performed well in comparison
with horizontal and vertical combining, with noticeably strong performance
in terms of the accuracy of the tails of the distributions. It was
also found to be competitive when compared against a variety of combining
and recalibration methods from the literature. We included an angular
combining version of the trimming-based method of \citet{Jose2014},
demonstrating that angular combining should not be viewed as a single
method because it allows a different approach to existing combining
methods.

\begin{SingleSpacedXII}{\small\bibliographystyle{apalike}
\bibliography{reference_all}
}\end{SingleSpacedXII}
\end{document}